\documentclass[12pt]{iopart}
\usepackage{appendix}
\usepackage{braket}
\usepackage{standalone}
\usepackage{graphicx}
\usepackage{iopams}
\usepackage{float}
\usepackage{subcaption}
\newcommand{\x}{(\bi{r}, t)}
\renewcommand{\Im}{\mathrm{Im}}
\renewcommand{\Re}{\mathrm{Re}}
\newcommand{\psit}{\ket{\Psi(t)}}
\newcommand{\dt}{\frac{d}{dt}}

\usepackage{cite}
\usepackage[colorlinks, citecolor = blue, urlcolor = blue]{hyperref}
\begin{document}

\title[]{Machine-learning Kohn-Sham Potential from Dynamics in Time-Dependent Kohn-Sham Systems}
\author{Jun Yang$^{1}$ \& James Whitfield$^{1,*}$}
\address{
$1$
Department of Physics and Astronomy, Dartmouth College. Hanover, NH 03755 
}

\address{$^{*}$
         Corresponding author:}
\ead{\mailto{james.d.whitfield@dartmouth.edu}}
\begin{abstract}
The construction of a better exchange-correlation potential in time-dependent density functional theory (TDDFT) can improve the accuracy of TDDFT calculations and provide more accurate predictions of the properties of many-electron systems. Here, we propose a machine learning method to develop the energy functional and the Kohn-Sham potential of a time-dependent Kohn-Sham system is proposed. The method is based on the dynamics of the Kohn-Sham system and does not require any data on the exact Kohn-Sham potential for training the model.  We demonstrate the results of our method with a 1D harmonic oscillator example and a 1D two-electron example. We show that the machine-learned Kohn-Sham potential matches the exact Kohn-Sham potential in the absence of memory effect. Our method can still capture the dynamics of the Kohn-Sham system in the presence of memory effects. The machine learning method developed in this article provides insight into making better approximations of the energy functional and the Kohn-Sham potential in the time-dependent Kohn-Sham system.
\end{abstract}

\maketitle
\section{Introduction}
A major challenge in TDDFT is the development of high-accuracy approximations for the Kohn-Sham potential $v_{KS}\x$ in the time-dependent Kohn-Sham (TDKS) system. The Kohn-Sham potential is an effective potential that allows us to equivalently convert the interacting electron system to a fictitious non-interacting system known as the Kohn-Sham system. As a tradeoff of this conversion, it is extremely hard to derive the explicit expression of the Kohn-Sham potential, which, in principle, determines the property of the electron system. As a result, significant research effort has been devoted to developing better approximations for the Kohn-Sham potential.

The Runge-Gross theorem\cite{Runge1984Mar} guarantees that the Kohn-Sham potential is uniquely determined (up to a purely time-dependent function) by the time-dependent electronic density $n(\bi{r}, t)$ for a given initial many-body state $\Psi_0 = \Psi(\bi{r}, t_0)$ of the interacting system and initial Kohn-Sham state $\Phi_0 =\Phi(\bi{r}, t_0)$ of the non-interacting Kohn-Sham system. The associated Kohn-Sham potential is often written as a sum of three terms $v_{KS}[n, \Psi_0, \Phi_0]\x = v_{ext}\x + v_H[n]\x + v_{xc}[n, \Psi_0, \Phi_0]\x$, where the first term is the external potential, the second term is the Hartree potential describing the classical electron-electron interaction. The first two terms are numerically easy and can be calculated explicitly given the electronic density. The last term is the exchange-correlation (xc) potential. The xc potential includes all the complex and non-trivial effects whose exact form is unknown.  Hence, the primary challenge in determining an accurate approximation of the Kohn-Sham potential is finding an effective approximation for the exchange-correlation potential. This task is challenging due to the complexity of electron interactions in many-electron systems.

To enhance the accuracy of TDDFT calculations and provide more precise predictions of the properties of many-electron systems, various approximation methods for the xc-potential have been developed. Commonly used methods include local density approximation (LDA), generalized gradient approximation (GGA), and meta-generalized gradient approximation (meta-GGA)\cite{Kohn1965Nov, Perdew1996Feb, Perdew1999Mar}, etc. Besides these traditional explicit constructions, some semi-empirical approaches were developed as well\cite{Lubasch2016Aug}. Recent works have provided approximation methods via machine learning techniques\cite{Nagai2018Jun, Suzuki2020May}. In Ref.~\cite{Suzuki2020May}, the authors investigated a two-electron system, where the electron densities and the exact xc-potential at different times were calculated explicitly from the exact form of the time-dependent two-electron wavefunction. A machine learning model that maps the electron density to the xc-potential was trained under the supervised learning framework, using the pre-calculated density-xc-potential pair as the dataset. Such a method only works in small systems because the computational cost of calculating the many-body wavefunction grows exponentially with the system size. Therefore, it is natural to ask whether we can develop a machine learning-based method for approximating the energy functional, and hence the Kohn-Sham potential and the xc-potential, without feeding the exact xc-potential to the machine learning model for training, since the xc-potential is uniquely determined by the electron density.

A similar question in classical mechanics has been extensively investigated by researchers from the artificial intelligence community. Several types of neural networks have been developed to learn the potential from classical trajectories. The well-trained model can be further utilized to predict the trajectories that are not contained in the training set. Most of the neural networks in learning classical dynamics can be classified into two categories, the Hamiltonian neural network (HNN)\cite{greydanus2019, Tong2021Jul, Bertalan2019Dec, Han2021May, Chen2019Sep} and the Lagrangian neural network (LNN)\cite{cranmer2020}. HNN incorporates Hamilton's equations into the neural network architecture, while LNN  uses ideas from the Lagrangian equation. Both methods have shown promising results for learning classical mechanics, but they have not yet been extended to learn the more challenging Kohn-Sham potential. 

Our work generalizes the idea of HNN to the quantum chemistry regime. We propose a machine learning method for studying the Kohn-Sham system under adiabatic approximation, and we provide a positive answer to the question we posed earlier about the feasibility of using machine learning to approximate the energy functional without using any information about the exact xc-potential. In section \ref{KS_HE}, we establish the connection between the Kohn-Sham equations and Hamilton’s equations. The connection lies as the cornerstone for the construction of our neural network. In section \ref{results}, we present our results with a harmonic oscillator example and a two-electron example. Conclusions and summaries are offered in section \ref{conclusion}.

\section{Kohn-Sham equations and Hamilton’s equations}\label{KS_HE}

Before discussing the Kohn-Sham equations, it is first necessary to understand the connection between the time-dependent Schr\"{o}dinger equation in quantum mechanics and Hamilton's equations in classical mechanics. The time-dependent Schr\"{o}dinger equation governs how a quantum system evolves over time, while Hamilton's equations determine the dynamical behavior of a classical mechanical system in terms of its position, momentum, and energy.

The time-dependent Kohn-Sham equations are a form of the time-dependent Schr\"{o}dinger equation, describing non-interacting particles. In TDDFT, the requirement for the Kohn-Sham equations is that the density produced by the non-interacting system should be the same as the density of the original interacting system\cite{Runge1984Mar, vanLeeuwen1998Feb}.

\subsection{Schr\"{o}dinger equation and Hamilton's equations}\label{section:sehe}

The Schr\"{o}dinger equation reads (in atomic units throughout the article):
\begin{equation}
    i\frac{\partial}{\partial t} \ket{\Psi(t)} = \hat{H} \ket{\Psi(t)}.
\end{equation}

Given state $\ket{\Psi(t)}$, we can rewrite it as a linear combination of a set of basis states $\{\ket{k}_{k=1,2,\ldots}\}$ (which leads to the set of basis functions in terms of position $s_k(\bi{r}) = \braket{\bi{r}|k}\}_{k=1,2,\ldots}$), expressed as $\ket{\Psi(t)} = \sum_k c_k(t) \ket{k}$. This expression allows us to rewrite the Schr\"{o}dinger equation into the form of a vector differential equation: 
\begin{equation}
    i\dot{\bi{c}}(t) = \bi{Hc}(t),
\end{equation} where $\bi{c}(t) = [c_1(t), \ldots, c_n(t), \ldots]^T$ is the coefficients vector, $\bi{H}$ is the matrix representation of the Hamiltonian $\hat{H}$ with the matrix elements being $H_{ij} = \braket{i|\hat{H}|j}$. To uncover the relationship between the Schr\"{o}dinger equation and Hamilton's equations, we define an energy functional:
\begin{equation}
    H[\bi{c}(t)] = \braket{\Psi(t)|\hat{H}|\Psi(t)} =  \bi{c^\dagger(t)} \bi{H}\bi{c}(t),
    \label{Eq:energy_se}
\end{equation}  where $\bi{c}^\dagger(t)$ is the complex conjugate of $\bi{c}(t)$.
 To recover Hamilton's equations in the form of real-valued differential equations, we can separate the coefficients vector into a real part and an imaginary part:
\begin{eqnarray}
    \bi{c}(t) = \frac{1}{\sqrt{2}}\left(\bi{q}(t) + i\bi{p}(t)\right),
    \label{Eq:complex_real}
\end{eqnarray}where $\bi{q}(t) =\sqrt{2}\, \Re\, \bi{c}(t)$ and $\bi{p}(t) = \sqrt{2}\,\Im\, \bi{c}(t)$.

This transformation in quantum mechanics was first proposed in Ref.~\cite{Strocchi1966Jan} and later investigated in Ref.~\cite{Kay1990Oct, Colbert1992Feb,GomezPueyo2018Jun}, allowing us to express the complex-valued energy functional $H[\bi{c}(t)]$ to real-valued functional $H[\bi{q}(t),\bi{p}(t)]$. Thus we have the following Hamilton's equations with respect to the real and imaginary part of the coefficients:
\begin{eqnarray}
    &\frac{\partial H[\bi{q}(t),\bi{p}(t)]}{\partial \bi{p}(t)} = \dot{\bi{q}}(t), \\
    &\frac{\partial H[\bi{q}(t),\bi{p}(t)]}{\partial \bi{q}(t)} = -\dot{\bi{p}}(t).
    \label{Eq:HE_SE_real}
\end{eqnarray}

A more compact equivalent expression in terms of complex variables is given by:
\begin{equation}
    i\dot{\bi{c}}(t) = \frac{\partial H[\bi{c}(t)]}{\partial \bi{c}^\dagger(t)},
    \label{eq:HE_SE}
\end{equation}
This equation connects the Schr\"{o}dinger equation and Hamilton's equation. In the next section, we will see how Kohn-Sham equations are connected to Hamilton's equations with a similar treatment.

\subsection{Kohn-Sham equations and Hamilton's equation}\label{section:KS_HE}
To derive similar equations in the time-dependent Kohn-Sham system, we need to go a bit beyond the Schr\"{o}dinger equation in the above section and design a different energy functional expression. To simplify things, we assume the adiabatic approximation $v_{xc}[n]\x = v_{xc}[n(t)](\bi{r})$ in the whole article\cite{Marques2004Apr, Casida2012Apr, Li2020Sep}, which is commonly used in TDDFT calculation. The adiabatic approximation means the exchange-correlation potential depends only on the instantaneous electron density $n(t)$ \footnote{The instantaneous electron density is also a function of position $n(\bi{r}, t)$. For the rest sections of the article, we use $n(t)$ to represent $n(\bi{r}, t)$ for simplicity.}. There are two reasons for using the adiabatic approximation. One is that the energy functional under adiabatic approximation is a functional of the instantaneous electron density, as a consequence, the xc-potential can be calculated from the xc-energy functional by $v_{xc}[n(t)](\bi{r}) = \frac{\delta E_{xc}[n(t)]}{\delta n(t)}$. The other consideration is that it is more convenient to construct a neural network that maps the instantaneous electron density to the energy functional.

Although the adiabatic approximation is widely used in TDDFT calculation, it often underestimates the complexity of the xc-potential. The exact xc-potential at time $t$ is typically dependent on the electron density at all previous times, leading to the so-called memory effect\cite{Liao2017Jun, Maitra2001Mar, Maitra2002Jun}. It is worth noting that such a complicated dependence on the history of the electron density is neglected by the adiabatic approximation. Thus, for a system where the memory effect is not negligible, our machine-learning method may not provide the exact Kohn-Sham potential. In the results part of this article, we will demonstrate this with examples.

With the adiabatic approximation, we can now discuss the relationship between the Kohn-Sham equations and Hamilton's equations. Note that the Kohn-Sham equations in TDDFT are time-dependent Schr\"{o}dinger equations describing the non-interacting particles:
\begin{eqnarray}
    i\frac{\partial }{\partial t} \ket{\phi_m(t)} = \hat{H}_{KS}[n(t)] \ket{\phi_m(t)}, m = 1, 2, \ldots,
\end{eqnarray} where $n(t) = \sum_m |\phi_m(t)|^2$, $\hat{H}_{KS}[n(t)] = -\frac{\nabla^2}{2} + v_{ext}\x + v_H[n(t)](\bi{r}) + v_{xc}[n(t)](\bi{r})$ is the Hamiltonian in the Kohn-Sham system,  $m$ labels the $m$-th Kohn-Sham orbital. The corresponding energy functional is given by:
\begin{equation}
    H_{KS}[n(t)] = T_s[n(t)]+ E_H[n(t)] + E_{ext}[n(t)] + E_{xc}[n(t)],
\end{equation} where $T_s[n(t)] = \sum_i \braket{\phi_i(t)|-\frac{\nabla^2}{2}|\phi_i(t)}$ is the kinetic energy functional in Kohn-Sham system, $E_H[n(t)] = \frac{1}{2}\int d\bi{r} d\bi{r}^\prime \frac{n(\bi{r}, t) n(\bi{r}^\prime, t)}{|\bi{r} - \bi{r}^\prime|}$, $E_{ext}[n(t)] = \int d\bi{r} v_{ext}(\bi{r})n(\bi{r},t)$ and $E_{xc}[n(t)]$ are the Hartree energy functional, external energy functional, and xc-energy functional, respectively.

Similar to the discussion in section~\ref{section:sehe}, it is possible to rewrite the Kohn-Sham equations into vector differential equations in terms of the coefficients of the Kohn-Sham orbitals given a set of basis funtions $\{s_k(\bi{r}) = \braket{\bi{r}|k}\}$:
\begin{equation}
    i\dot{\bi{c}}_m(t) = \bi{H}_{KS}[n(t)]\bi{c}_m(t),
\end{equation} where $\bi{c}_m(t) = [\braket{1|\phi_m(t)}, \braket{2|\phi_m(t)}, \dots, \braket{n|\phi_m(t)}, \ldots]^T$ is the coefficient vector of the $m$-th Kohn-Sham orbital, and the matrix element of $\bi{H}_{KS}[n(t)]$ is given by $\braket{i|\hat{H}_{KS}[n(t)]|j}$. Since the electron density can be expressed in terms of the coefficient vector $n(t) = \sum_{m,i,j} c_{mi}(t)c_{mj}^*(t)s_i(\bi{r})s_j^*(\bi{r})$, we can use $\bi{H}_{KS}[\{\bi{c}_m(t)\}_{m = 1, 2\ldots}]$ and $H_{KS}[\{\bi{c}_m(t)\}_{m = 1, 2\ldots}]$ to replace $\bi{H}_{KS}[n(t)]$ and $H_{KS}[n(t)]$, respectively. Thus the total energy functional defined in the Kohn-Sham system is consistent with the energy functional defined in Eq.~\ref{Eq:energy_se}

Similarly, we can show that Hamilton's equations hold in the Kohn-Sham system under the adiabatic approximation as well. The details of the proof are shown in \ref{pf:ks_he}.
\begin{equation}
     i \dot{\bi{c}}_m(t) = \frac{\partial H_{KS}[\{\bi{c}_m(t)\}_{m = 1, 2\ldots}]}{\partial \bi{c}^\dagger_m(t)}, \,m = 1, 2, \ldots
     \label{eq:HE_KS}
\end{equation} An equivalent real-valued Hamilton's equation can be obtained by applying the transformation given in Eq.~\ref{Eq:complex_real}:
\begin{eqnarray}
     &\dot{\bi{q}}_m(t) = \frac{\partial H_{KS}[\{\bi{q}_m(t), \bi{p}_m(t)\}_{m = 1, 2\ldots}]}{\partial \bi{p}_m(t)}, \\
     &\dot{\bi{p}}_m(t) = -\frac{\partial H_{KS}[\{\bi{q}_m(t), \bi{p}_m(t)\}_{m = 1, 2\ldots}]}{\partial \bi{q}_m(t)}, \,m = 1, 2, \ldots
\end{eqnarray}

\begin{figure}[!htbp]
    \centering
    \includegraphics[width=\textwidth]{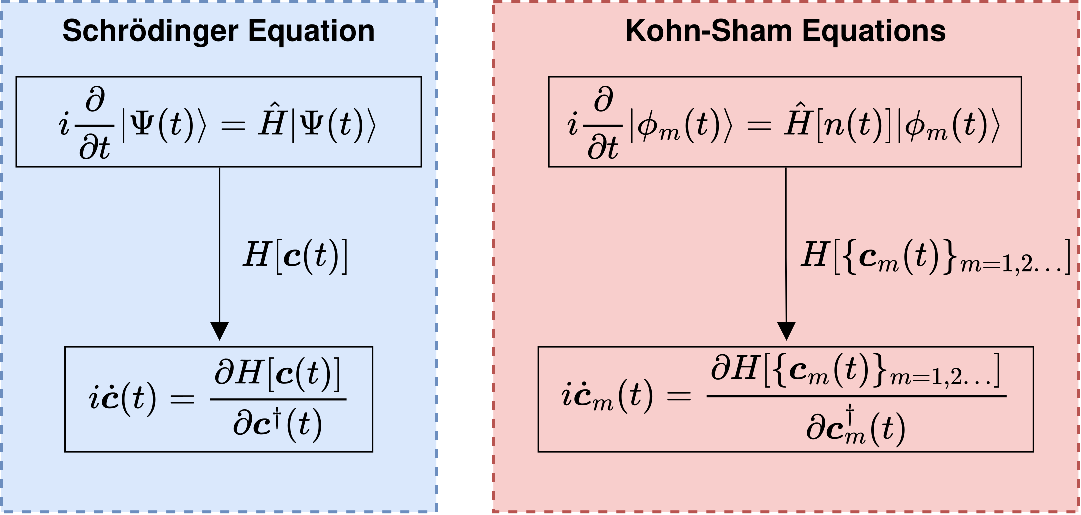}
    \caption{Workflow for converting quantum equations into classical Hamilton's equation. The left blue block is the workflow for Schr\"{o}dinger equation. The right red block is the workflow for Kohn-Sham equations.}
    \label{fig:SEKS}
\end{figure}

Therefore, Eq.~\ref{Eq:HE_SE_real} and \ref{eq:HE_SE} can be generalized to the time-dependent Kohn-Sham system. In Fig.~\ref{fig:SEKS}, we show the workflow for converting the quantum equations to classical Hamilton's equations in both the Schr\"{o}dinger equation and the Kohn-Sham equations. 
The Hamilton's equations derived in the Kohn-Sham system are the cornerstone of our machine-learning method. We will discuss the details in the next section.

\subsection{Machine learning energy functional}

From this part to the end, the discussion is limited to the 1-dimensional system. For simplicity, we use $H[\bi{c}(t)]$ and $H[\bi{q}(t),\bi{p}(t)]$ instead of $H_{KS}[\{\bi{c}_m(t)\}_{m = 1, 2\ldots}]$ and $H_{KS}[\{\bi{q}_m(t), \bi{p}_m(t)\}]$ for the following discussion, because as shown in Eq.~\ref{eq:HE_KS}, all Kohn-Sham orbitals follow the same differential equation. In this section, we will demonstrate how we incorporate the equations derived above into the neural networks.

Before discussing the technical details, we need to prepare the dataset. We then pick $n$ basis functions. In our work, the sinc discrete variable representation (DVR) basis $\braket{x|k} = \sqrt{\Delta x} \frac{\mathrm{sinc}\left[\frac{\pi(x - k\Delta x)}{\Delta x}\right]}{\pi(x - k\Delta x)}$ is used\cite{Colbert1992Feb, Brown2020Oct}, where $\Delta x$ is the grid's spacing. Thus for any given initial condition $\bi{c}(t = 0) = \bi{c}_0$, $\bi{c}(t) = [c_1(t), c_2(t),\ldots, c_n(t)]$ defines a trajectory $\gamma(t)$ in $\mathbb{C}^n$, or equivalently $\mathbb{R}^{2n}$. In the rest of the discussion, we will use the real-valued Hamiltonian since the real-valued functions are better handled in contemporary machine learning frameworks, e.g., PyTorch\cite{NEURIPS2019_9015}, TensorFlow\cite{Abadi2016Mar}, etc. Given $M$ different initial conditions, we can obtain $M$ different trajectories $\gamma^{(1)}(t), \dots, \gamma^{(M)}(t)$. We can then perform sampling on the trajectories over $N$ different timestamps. The total $M \times N$ sampling points and their time-derivatives $\{(\gamma^{(1)}(t_0), \dot{\gamma}^{(1)}(t_0)), \ldots, (\gamma^{(M)}(t_N), \dot{\gamma}^{(M)}(t_N))\}$ make up our dataset.

\begin{figure}
    \centering
    \includegraphics[width=\textwidth]{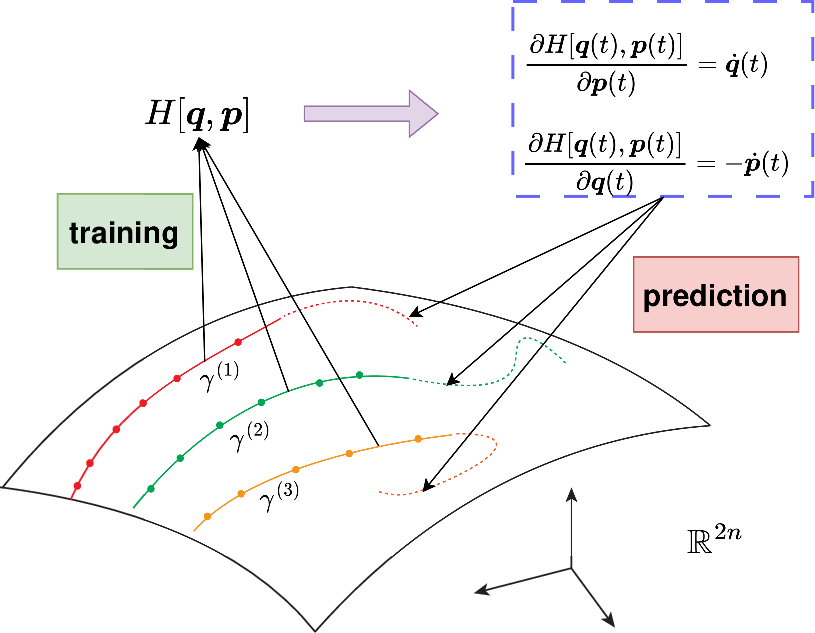}
    \caption{Illustration of the Machine learning Hamiltonian framework. The sampling points over $M$ trajectories are used for training a Hamiltonian modeled by a neural network, which will be used to predict the dynamics of the system.}
    \label{fig:Ham_traj}
\end{figure}

The Hamiltonian $H[\bi{q}, \bi{p}]$ defines a real-valued function on $\mathbb{R}^{2n}$, which determines one trajectory given an initial condition. Thus we can use a neural network ansatz to parameterize the Hamiltonian that controls the system's dynamics and train it with the dataset we obtained above. The trained Hamiltonian can be used to predict the future dynamics of the system. This idea is illustrated in Fig.~\ref{fig:Ham_traj}. 
An important property of this framework is that the total energy is invariant on a given trajectory $\gamma(t)$ because 
\begin{eqnarray}
    \frac{d}{dt} H[\bi{q}(t), \bi{p}(t)] &= \frac{\partial H[\bi{q}(t),\bi{p}(t)]}{\partial \bi{p}(t)} \dot{\bi{p}}(t) + \frac{\partial H[\bi{q}(t),\bi{p}(t)]}{\partial \bi{q}(t)}\dot{\bi{q}}(t)=0
    \label{Eq:conservation}
\end{eqnarray}

\begin{figure}[!htbp]
        \centering
\includegraphics[width=\textwidth]{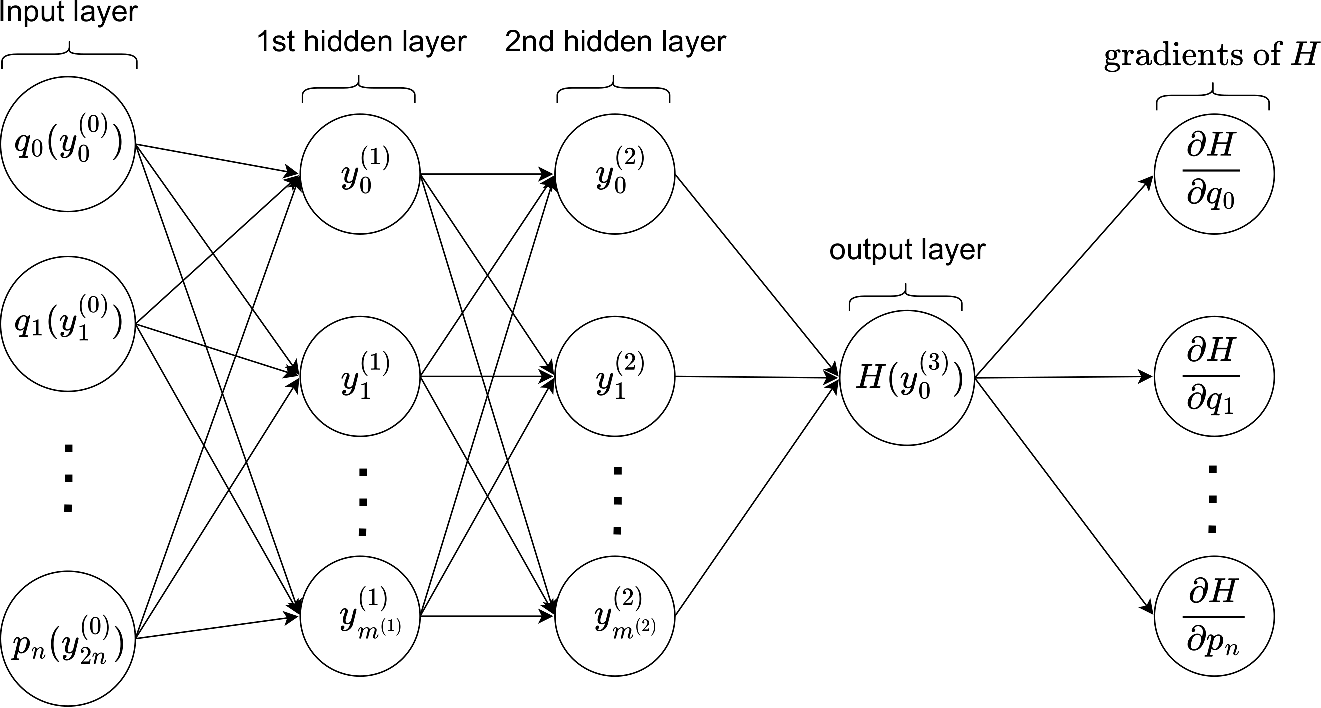}
        \caption{The structure of our neural network. $q_i = \sqrt{2} \Re \,c_i$ is the scaled real part of the $i$-th entry of the coefficient vector, $p_i = \sqrt{2} \Im \,c_i$ is the scaled imaginary part of the $i$-th entry of the coefficient vector. The scaled real and imaginary parts of the coefficients at time $t$ are fed into the neural network as our inputs. There are two hidden layers. Neurons are connected by non-linear activation functions (in the numerical experiments we carried out, softplus and tanh were chosen). The output layer is the energy functional of the Kohn-Sham system. The loss function relies on the gradient (shown in the last layer in the figure) of the energy functional, we compute it with auto differentiation functionality in Pytorch.}
        \label{fig:nn_structure}
    \end{figure}

The structure of our neural network is shown in Fig.~\ref{fig:nn_structure}. The input layer of the neural network has 2$n$ entries, the first $n$ entries are for the scaled real part of the coefficients at time $t$, and the last $n$ entries are for the scaled imaginary part of the coefficients. There are two hidden layers. There is one scalar as the output layer representing the energy functional of the system in our neural network. Let $\bi{y}^{(i)}$ be the variable vector of the $i$-th layer, the $i$-th and the $i+1$-th layer are connected by,
\begin{equation}
    \bi{y}^{(i + 1)} = \sigma(\bi{W}^{(i)}\bi{y}^{(i)} + \bi{b}^{(i)}),
\end{equation} where $\sigma(\cdot)$ is the non-linear activation function, $\bi{W}^{(i)}$ and $\bi{b}^{(i)}$ are trainable weights and biases of the $i$-th layer.  In the numerical experiments we carried out, $\sigma(x) = \tanh{x}$ and $\sigma(x) = \mathrm{softplus}(x)$ were used for the activation functions.

The loss function is given by: 
\begin{equation}
    \mathcal{L}=\frac{1}{n}\sum_{i=1}^n \left(\left|\dot{q}_i - \frac{\partial H}{\partial p_i}\right|^2 + \left|\dot{p}_i + \frac{\partial H}{\partial q_i}\right|^2\right),
\end{equation}where $q_i = \sqrt{2} \Re \,c_i$ and $p_i = \sqrt{2} \Im\, c_i$ are the scaled real and imaginary parts of the $i$-th entry of the coefficient vector $\bi{c}$, respectively. $\dot{q}_i$ and $\dot{p}_i$ are their time derivatives. $\frac{\partial H}{\partial q_i}$ and $\frac{\partial H}{\partial p_i}$ are from the gradient of the energy functional calculated using the auto-differentiation functionality in modern machine learning frameworks. In our work, we chose Pytorch.  The Hamilton's-equation-inspired loss function was originally proposed in Ref.~\cite{DeWilde1993Feb} and Ref.~\cite{greydanus2019} for studying classical motions. We generalize it to the quantum regime for investigating potential inversion problems in Schr\"{o}dinger equations and Kohn-sham equations.  The neural network is optimized using the adaptive moment estimation method (Adam)\cite{Kingma2014Dec}. By minimizing the loss function, we can get the energy functional of the Kohn-Sham system parametrized by the neural network. 

We need to take an additional step to calculate the Kohn-Sham potential from the energy functional in the system. Since the sinc DVR is used in this article, the corresponding matrix element of the potential term $V_{KS}^{ij}$ is given by:
\begin{equation}
    V_{KS}^{ij} = V_{KS}(x_i)\delta^{ij} = \frac{1}{2} \left(\frac{\partial^2 H[\bi{q},\bi{p}]}{\partial q_i \partial q_j}+ \frac{\partial^2 H[\bi{q},\bi{p}]}{\partial p_i \partial p_j}\right) - T^{ij},
    \label{eq:v_ks}
\end{equation}
where $T^{ij}$ is the matrix element of the kinetic energy term whose value is\cite{Colbert1992Feb, Brown2020Oct} 
\begin{equation}
    T^{ij} = \frac{(-1)^{(i-j)}}{2\Delta x^2}
    \cases{\frac{\pi^2}{3}, & $i = j$ \\ 
       \frac{2}{(i-j)^2}, & $i \neq j$.}
\end{equation}
The details of the proof of Eq.~\ref{eq:v_ks} are shown in \ref{pf:ks_he}.
\section{Results}\label{results}
\subsection{Harmonic oscillator}
The first example we choose is the 1-dimensional harmonic oscillator. With this example, we will demonstrate that the system dynamics can be successfully reproduced by the abovementioned machine learning method. 

The potential is time-independent $v(x) = \frac{1}{2}\omega^2 x^2$, where $\omega = 1$. 
The eigenstates and their time derivatives of the Harmonic oscillator have analytical expressions,
\begin{eqnarray}
    &\varphi_{n}(x, t)={\frac {1}{\sqrt {2^{n}\,n!}}}\left({\frac {1}{\pi }}\right)^{1/4}e^{-{\frac {x^{2}}{2}} -i(n + \frac{1}{2})t}H_{n}\left(x\right), \\
    &\dot{\varphi}_{n}(x, t)={\frac {-i(n + \frac{1}{2})}{\sqrt {2^{n}\,n!}}}\left({\frac {1}{\pi }}\right)^{1/4}e^{-{\frac {x^{2}}{2}} -i(n + \frac{1}{2})t}H_{n}\left(x\right),
    \\
    &\nonumber n=0,1,2,\ldots, M-1,
\end{eqnarray} where $H_n(x)$ is the $n$-th Hermite polynomial. We obtain the dataset from the lowest $M$ eigenstates. For each eigenstate in the dataset, we chose 150 evenly spaced grid points in the range $x\in [-6, 6]$. In the time domain, 2000 timestamps are evenly sampled in $t\in[0, 4\pi]$ with a time step length $\Delta t = 6.3 \times 10^{-3}\,(a.u.)$. Thus, the input dimension of the neural network is 300 (twice the number of grid points), and the dimension of each hidden layer is 400. The number of eigenstates is set to be $M=15$ in this numerical experiment, which means we have sampled over 15 different trajectories in this dataset, each of which contains 2000 sampling points.

\begin{figure}[!htbp]
\centering
    \includegraphics[width=\textwidth]{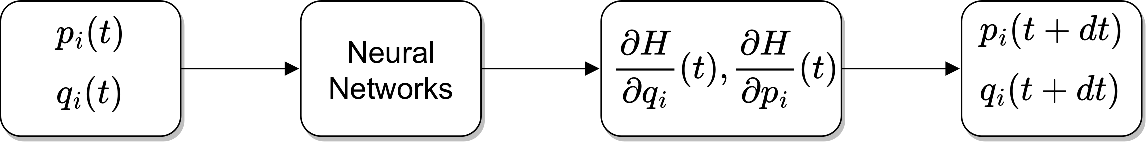}
\caption{Pipeline of propagating the initial state with our neural network}
\label{Fig:pipline}
\end{figure}

 To verify the trained neural network produces the correct dynamics of the system, we randomly pick an initial state that is not in the training set and let it propagate under the given Hamiltonian. In the example given below, the initial state is a linear combination of the lowest four eigenstates $\ket{\psi(x, t = 0)} = \frac{1}{\sqrt{2}} \ket{0} + \frac{1}{2}\ket{1} + \frac{1}{\sqrt{6}} \ket{2} + \frac{1}{\sqrt{12}}\ket{3}$. Then we compare the machine-learned time-dependent density to the exact density. To be precise, at time $t$, we can compute the gradient of the energy functional $\frac{\partial H}{\partial q_i}$ and $\frac{\partial H}{\partial p_i}$ using auto-differentiation, which can be used to propagate the coefficients to the next timestamp $t + dt$ with the Runge-Kutta (R-K) method\cite{Runge1895}. In this sense, our neural network can be viewed as a differential equation solver. The entire pipeline of propagating the initial state with our neural network is summarized in Fig.~\ref{Fig:pipline}. 
 
\begin{figure}[!htbp]
\centering
\begin{subfigure}[b]{0.32\textwidth}
    \centering
    \includegraphics[width=\textwidth]{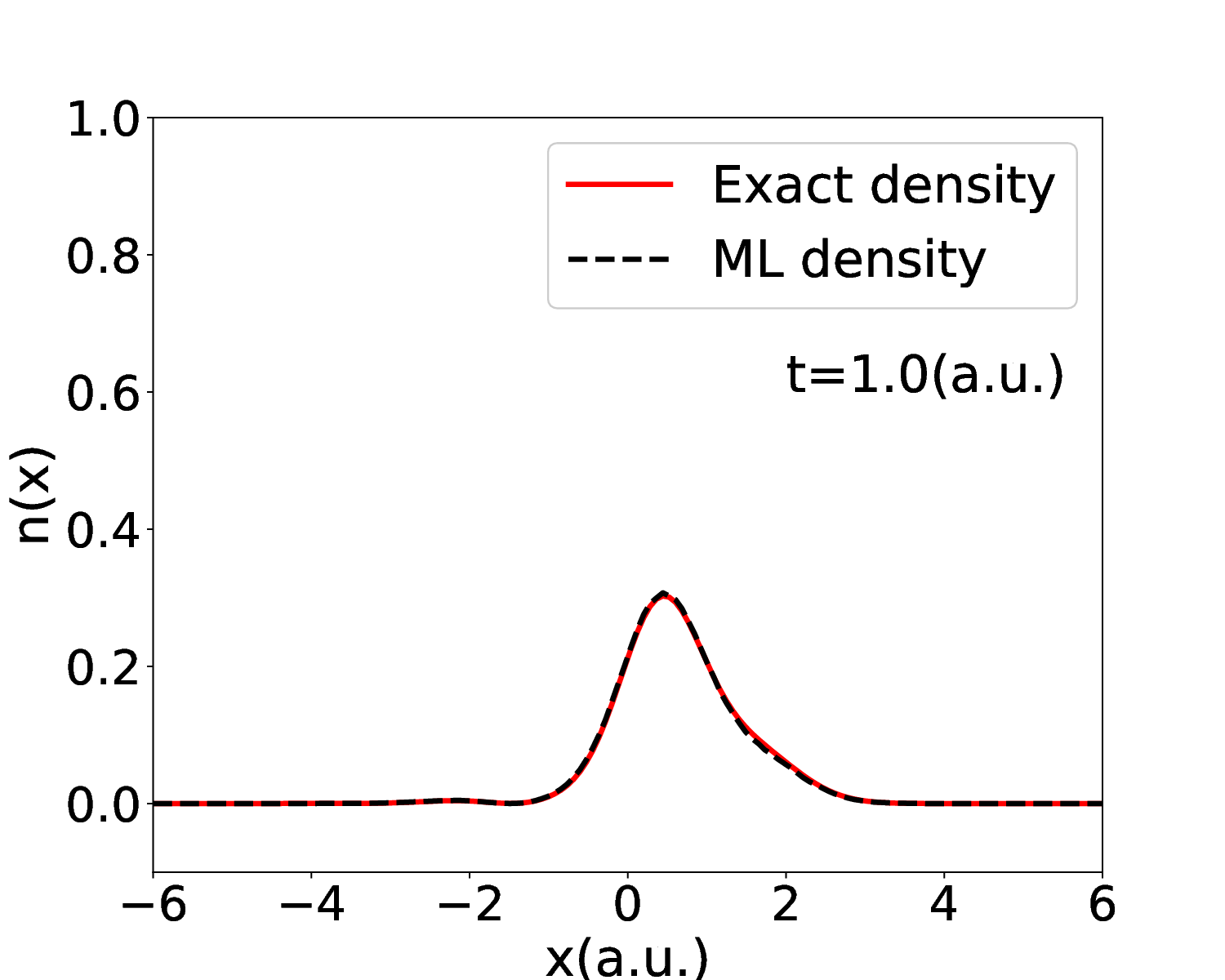}
\end{subfigure}
\begin{subfigure}[b]{0.32\textwidth}  
    \centering 
    \includegraphics[width=\textwidth]{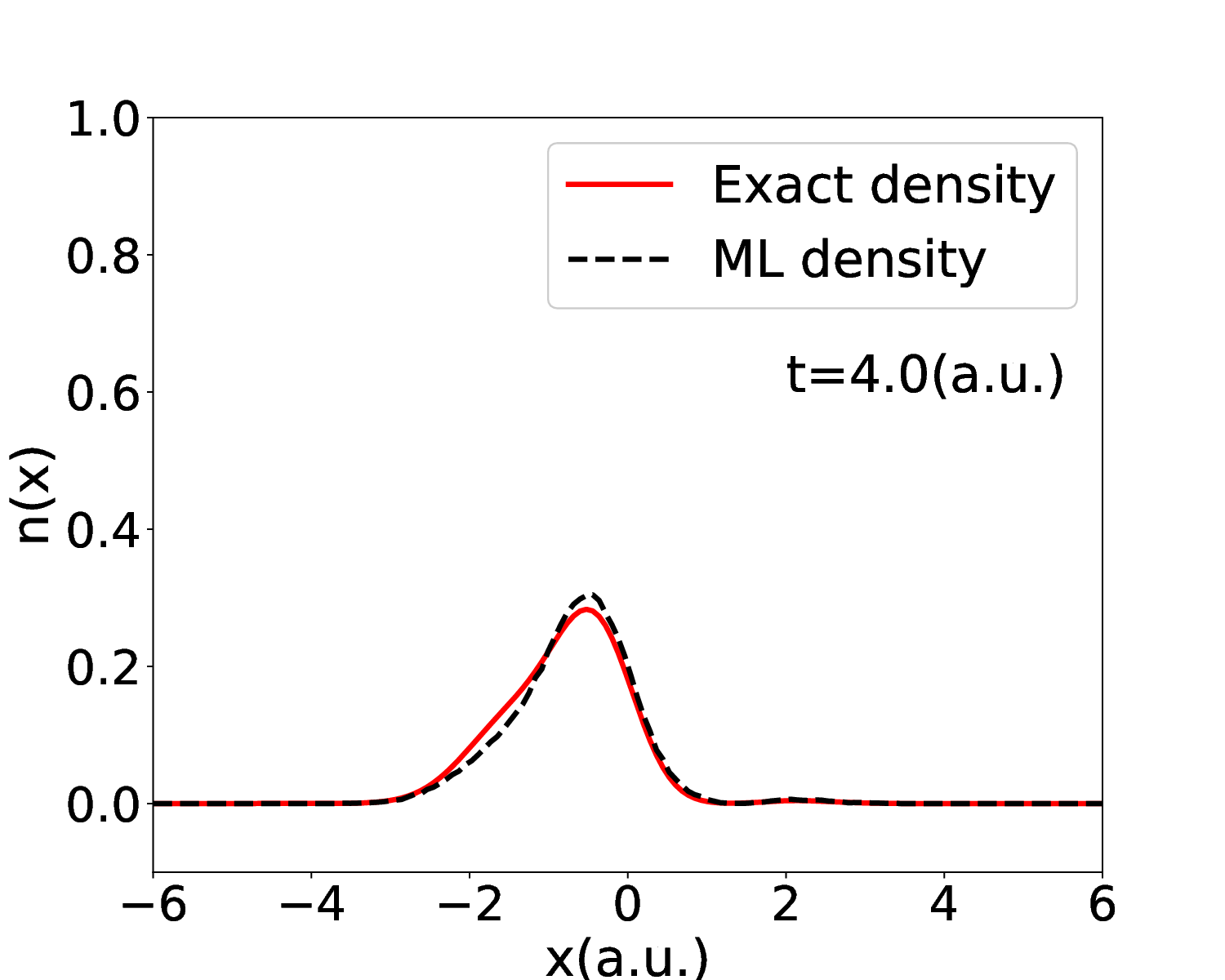}
\end{subfigure}
\begin{subfigure}[b]{0.32\textwidth}  
    \centering 
    \includegraphics[width=\textwidth]{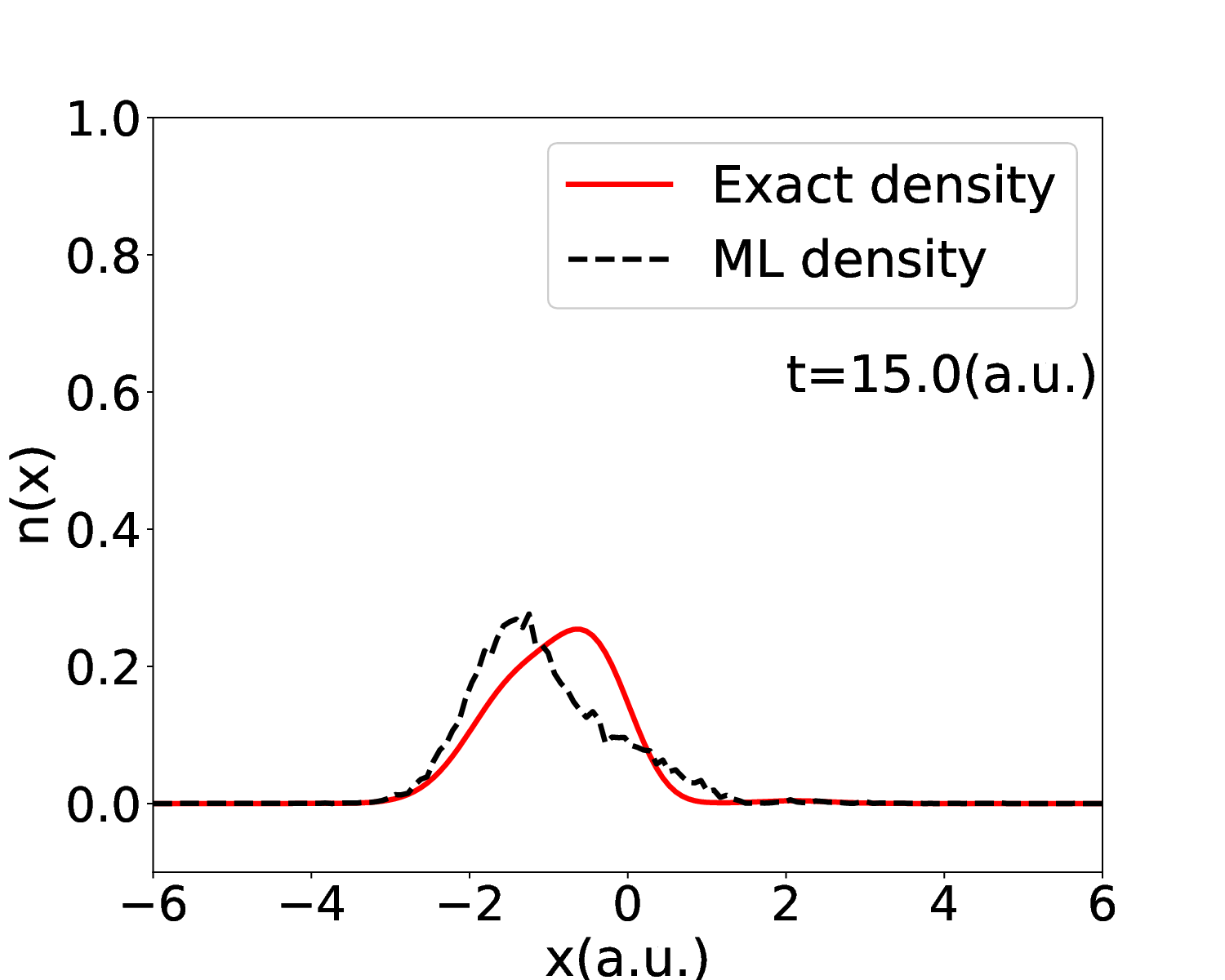}
\end{subfigure}
    \caption{The machine learned  density and the exact density at different timestamps. The black dashed line is the machine-learned density, the red solid line is the exact density. The three panels correspond to the results of timestamp $t = 1.0\, (a.u.)$ (left), $t = 4.0\, (a.u.)$ (middle), and $t = 15.0 \,(a.u.)$ (right).}
    \label{fig:td_density}
\end{figure}

 We compare the time evolution of the electron density propagated using the machine-learned energy functional and the exact result calculated from analytical expression in Fig.~\ref{fig:td_density}. The machine-learned density (black dashed line) exhibits a similar trend to the exact result at any point in the entire position space at early timestamps, e.g., $t = 1.0 \,(a.u.)$. The error accumulates with time. As shown in Fig.~\ref{fig:td_density}, the error grows larger at $t=4.0 \, (a.u.)$, but still captures the general feature of the exact density. In the third plot of Fig.~\ref{fig:td_density}, the results are completely different at a longer time $t = 15.0\, (a.u.)$.

\begin{figure}[!htbp]
\centering
\begin{subfigure}[b]{0.45\textwidth}
    \centering
    \includegraphics[width=\textwidth]{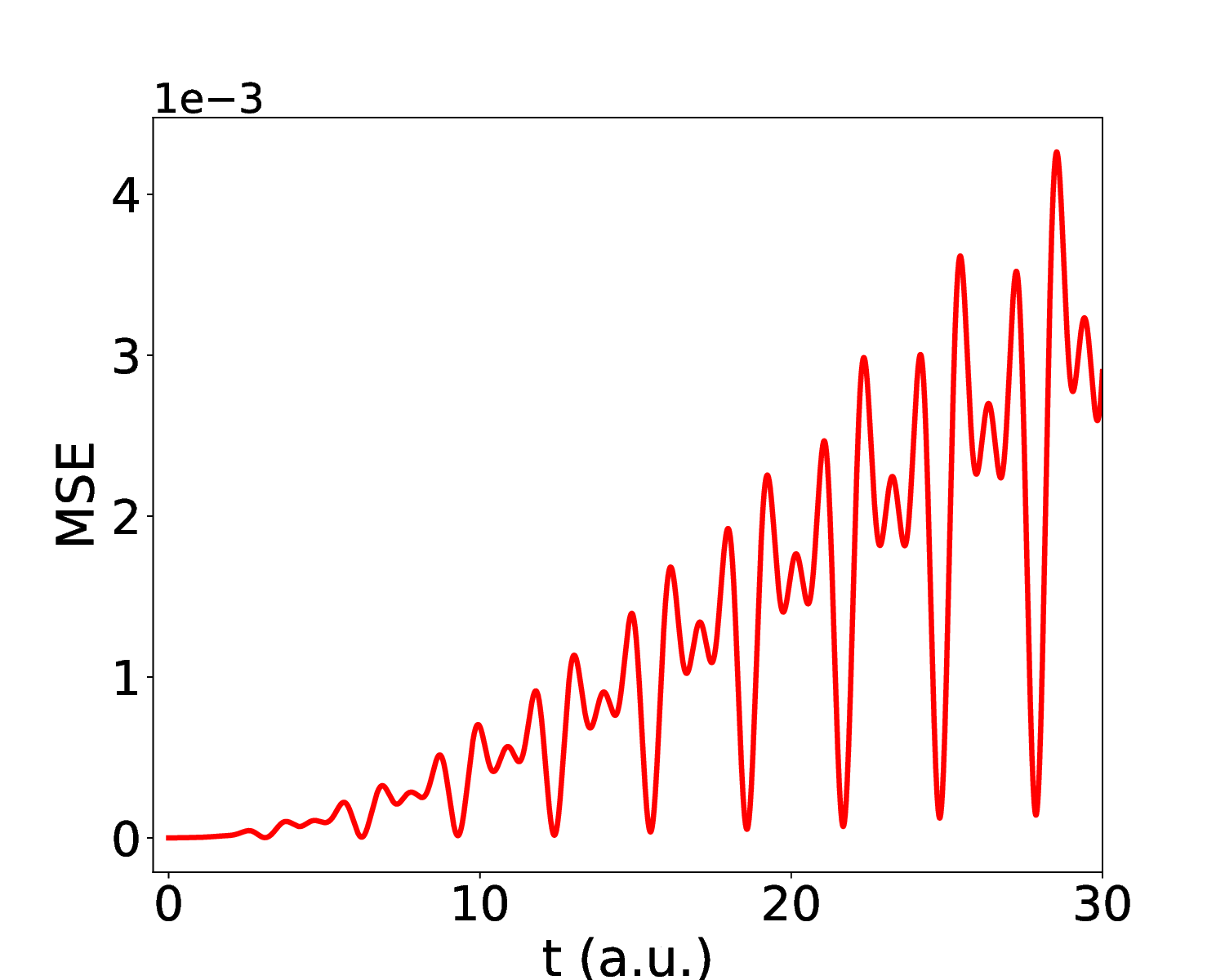}
    \caption{}
\end{subfigure}
\begin{subfigure}[b]{0.45\textwidth}
    \centering
    \includegraphics[width=\textwidth]{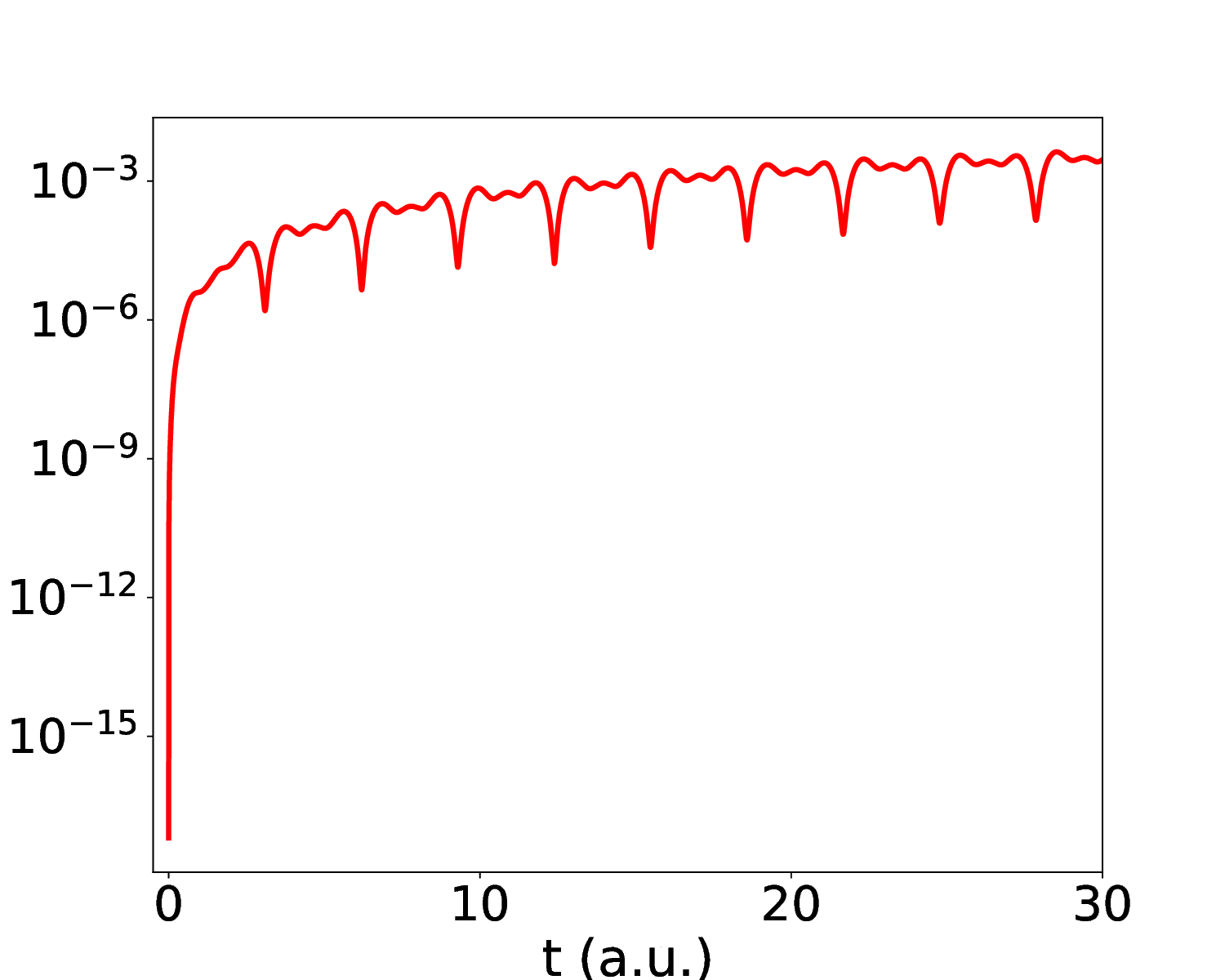}
    \caption{}
\end{subfigure}
    \caption{(a) MSE in density accumulates with time in linear scales. (b) MSE in density accumulates with time in logarithmic scales. The periodic pattern in local maximum and local minimum due to the periodicity of the system.}
    \label{fig:ho_err}
\end{figure}

We use mean square error (MSE) to quantify the error in the density at each timestamp. Given the exact electron density $n(x,t)$ and the machine-learned density $n_{ML}(x,t)$, the MSE at time $t$ between $n(x,t)$ and $n_{ML}(x,t)$ is given by $\frac{1}{N_g}\sum_{i=1}^{N_g} (n_{ML}(x_i,t) - n(x_i,t))^2$, where $N_g$ is the number of grid points in the space domain. The resulting MSE at different times $t$ and its logarithmic plot are shown in Fig.~\ref{fig:ho_err}. It is clear from the figure that the MSE grows in time. We notice the error in this example shows a periodic pattern; this is due to the periodicity of the system. To avoid the error from going too big, one can set a threshold and calibrate the time evolution of the trajectory when the error is larger than a predefined threshold. 

\begin{figure}[!htbp]
\centering
\begin{subfigure}[b]{0.4\textwidth}
    \centering
    \includegraphics[width=\textwidth]{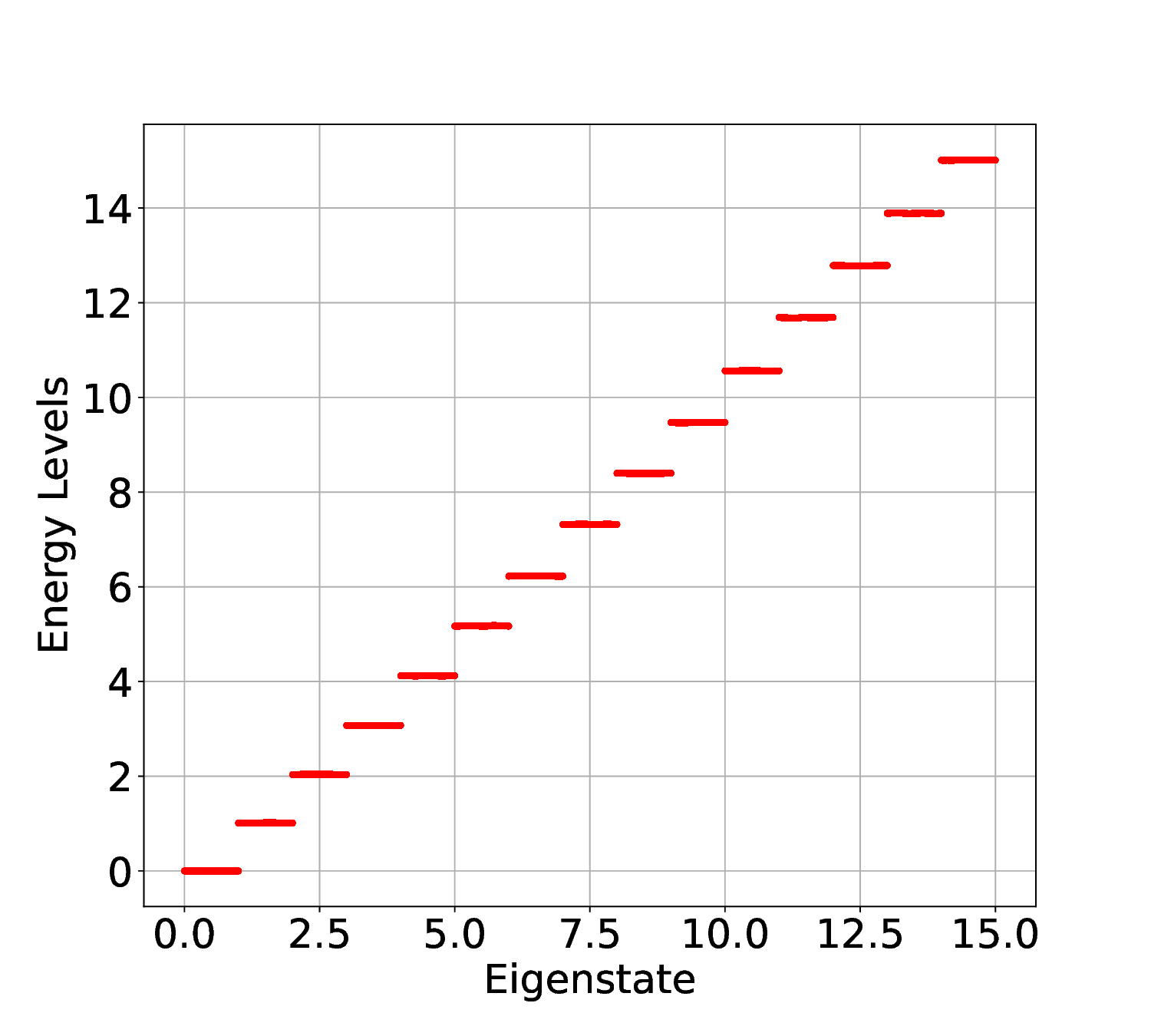}
    \caption{}
    \label{fig:energy_level}
\end{subfigure}
\begin{subfigure}[b]{0.4\textwidth}  
    \centering 
    \includegraphics[width=\textwidth]{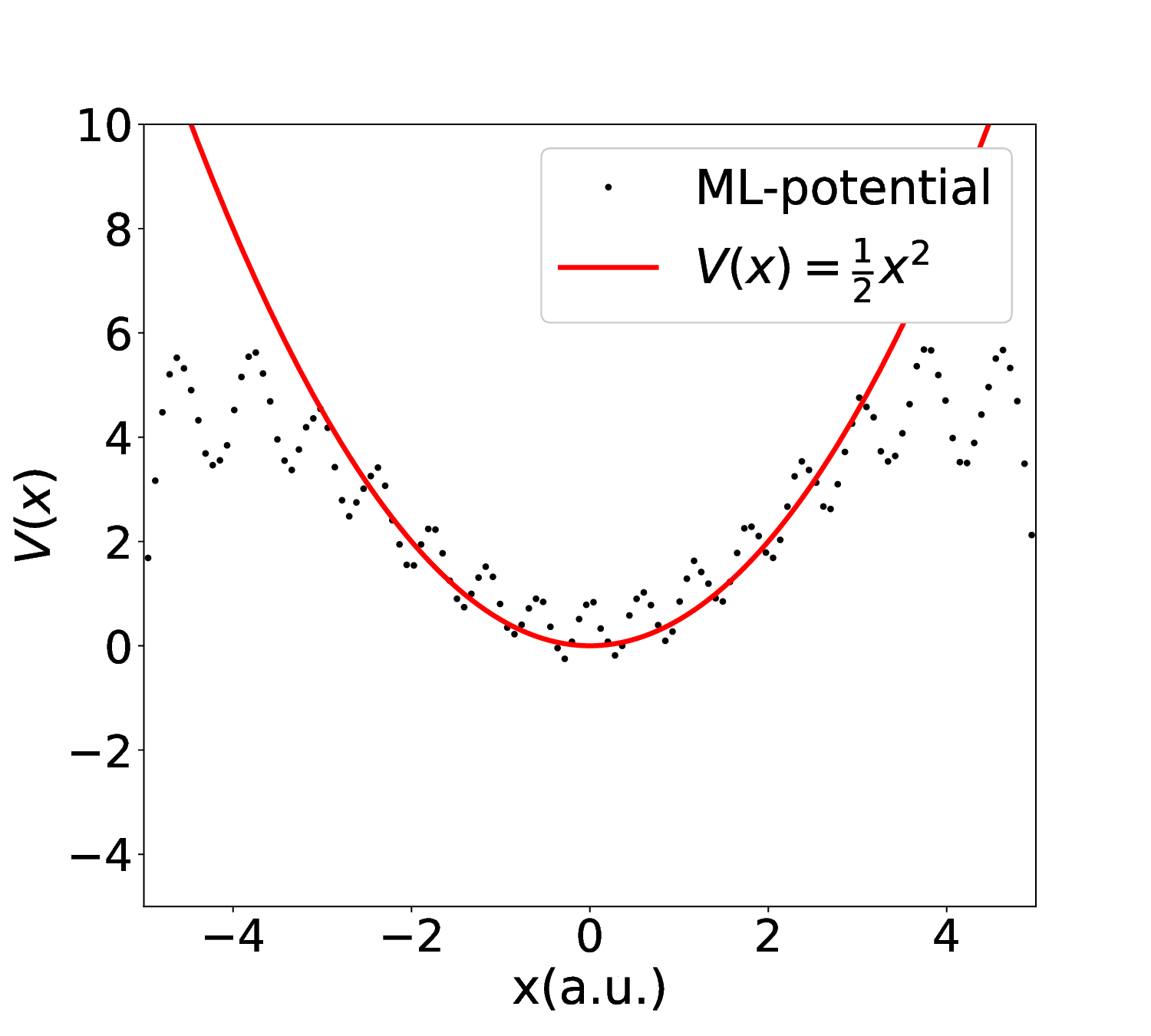}
    \caption{}
    \label{fig:ml_potential}
\end{subfigure}
    \caption{(a) is the figure of energy level. We manually shift the ground state energy to 0 for comparison. The neural network generates the correct level spacing for the harmonic oscillator test. (b) the machine learned potential (black dots) versus the actual potential (red line).}
\end{figure}

As shown in Eq.~\ref{Eq:conservation}, the energy functional on a given trajectory shouldn't be changed over time. Specifically, in our harmonic oscillator test, if we evaluate the machine-learned energy functional on the trajectory of an eigenstate $\phi_n(x, t)$, the output should be the eigenvalue $n + \frac{1}{2}$ (up to a constant shift). This phenomenon is observed in Fig.~\ref{fig:energy_level}, where each red stages from bottom to top correspond to the trajectories of the first to the 15th eigenstates, respectively. As shown in the figure, each stage is flat, indicating that the energy associated with the trajectory remains constant over time. Another interesting fact is that the difference between consecutive stages is 1, which is exactly the energy level spacing of the Harmonic oscillator. These are non-trivial effects because energy conservation law has never been explicitly involved in the whole machine learning workflow. However, the neural network can discover the pattern from data automatically.

We calculate the machine-learned potential using Eq.~\ref{eq:v_ks} and show the machine-learned potential (black dots) versus the exact potential (red line) in Fig.~\ref{fig:ml_potential}. The machine-learned potential captures the structure of the exact quadratic potential around the center of the position space, but it starts to deviate from the actual potential near the boundary of the pre-selected finite region. One possible reason for this phenomenon is that our training set only covers a fraction of the entire Hilbert space, which means (i) the training set covers only a finite range of the infinite position space, (ii) the training set includes only the lowest $M$ eigenstates of the infinite many eigenstates. We show in \ref{app: v_scale} that having more eigenstates in the training set results in more accurate machine-learned potential.

\subsection{Two electron test}
We revisit the two-electron system previously studied in\cite{Lacombe2018Jun,Suzuki2017Dec,Suzuki2020May}, and use the machine learning method developed in our work to study the Kohn-Sham potential in the system. With this example, we demonstrate that our machine-learning method can be extended to study realistic electronic systems as well.  The many-body Hamiltonian governing the dynamics of the two-electron system is given by:
\begin{equation}
    \hat{H}\left(x_{1}, x_{2}\right)=\sum_{i=1,2}\left[-\frac{1}{2} \frac{\partial^{2}}{\partial x_{i}^{2}}+v_{\mathrm{ext}}\left(x_{i}\right)\right]+W_{e e}\left(x_{1}, x_{2}\right), 
\end{equation} where $W_{ee}\left(x_{1}, x_{2}\right)=\frac{1}{\sqrt{\left(x_{1}-x_{2}\right)^{2}+1}},v_{\mathrm{ext}}=-\frac{1}{\sqrt{\left(x_{1}+10\right)^{2}+1}}$. The spatial part of the initial two-electron wavefunction is $\Psi(x_1, x_2) = \frac{1}{\sqrt{2}}\left[\phi_{\mathrm{H}}\left(x_{1}\right) \phi_{\mathrm{WP}}\left(x_{2}\right)+\phi_{\mathrm{WP}}\left(x_{1}\right) \phi_{\mathrm{H}}\left(x_{2}\right)\right]$, where $\phi_H(x)$ is the ground state hydrogen wavefunction, $\phi_{WP}(x) = (2 \alpha / \pi)^{\frac{1}{4}} e^{\left[-\alpha\left(x-x_{0}\right)^{2}+i p\left(x-x_{0}\right)\right]}$ with $\alpha = 0.1, x_0 = 10.0, p = -1.5$. So we have one Gaussian wave packet centered at $x_0 = 10.0$ moving leftward. As the system evolves, the Gaussian wave packet collides with the static Hydrogen wavefunction and then passes through the Hydrogen wave function at longer times.

The two-electron wavefunction $\Psi(x_1, x_2, t)$ can be computed numerically. Based on the numerically exact solution of the two-electron wavefunction, we can calculate the electronic density $n(x, t) = 2 \int dx_2 \Psi^\star(x, x_2, t) \Psi(x, x_2, t)$ and the density current $j(x, t) = 2 \Im \int dx_2 \Psi^\star(x,x_2,t)\partial_x \Psi(x, x_2, t)$. For a spin-singlet state, the spatial parts of the two Kohn-Sham orbitals are computed as $\phi_1(x,t) = \phi_2(x, t) = \sqrt{\frac{n(x, t)}{2}} e^{i \int^{x} \frac{j\left(x^{\prime}, t\right)}{n\left(x^{\prime}, t\right)} d x^{\prime}}$, such that the corresponding Kohn-Sham potential is a real function\cite{Elliott2012Dec, Maitra2002Jun}. Coefficients of the Kohn-Sham orbital are then obtained as our dataset.

Similar to the discussion in the harmonic oscillator example, the system is discretized evenly in the range $x\in[-24.78, 21.97]$ with 200 grid points. The propagation starts from $t = 0$ to $t = 15.0 \,(a.u.)$ with time step length $\Delta t = 5 \times 10^{-4} \,(a.u.)$. The input dimension of the neural network for the two-electron model is thus 400, and the dimension of each hidden layer is 400.

\begin{figure}[!htbp]
\centering
\begin{subfigure}[b]{0.32\textwidth}
    \centering
    \includegraphics[width=\textwidth]{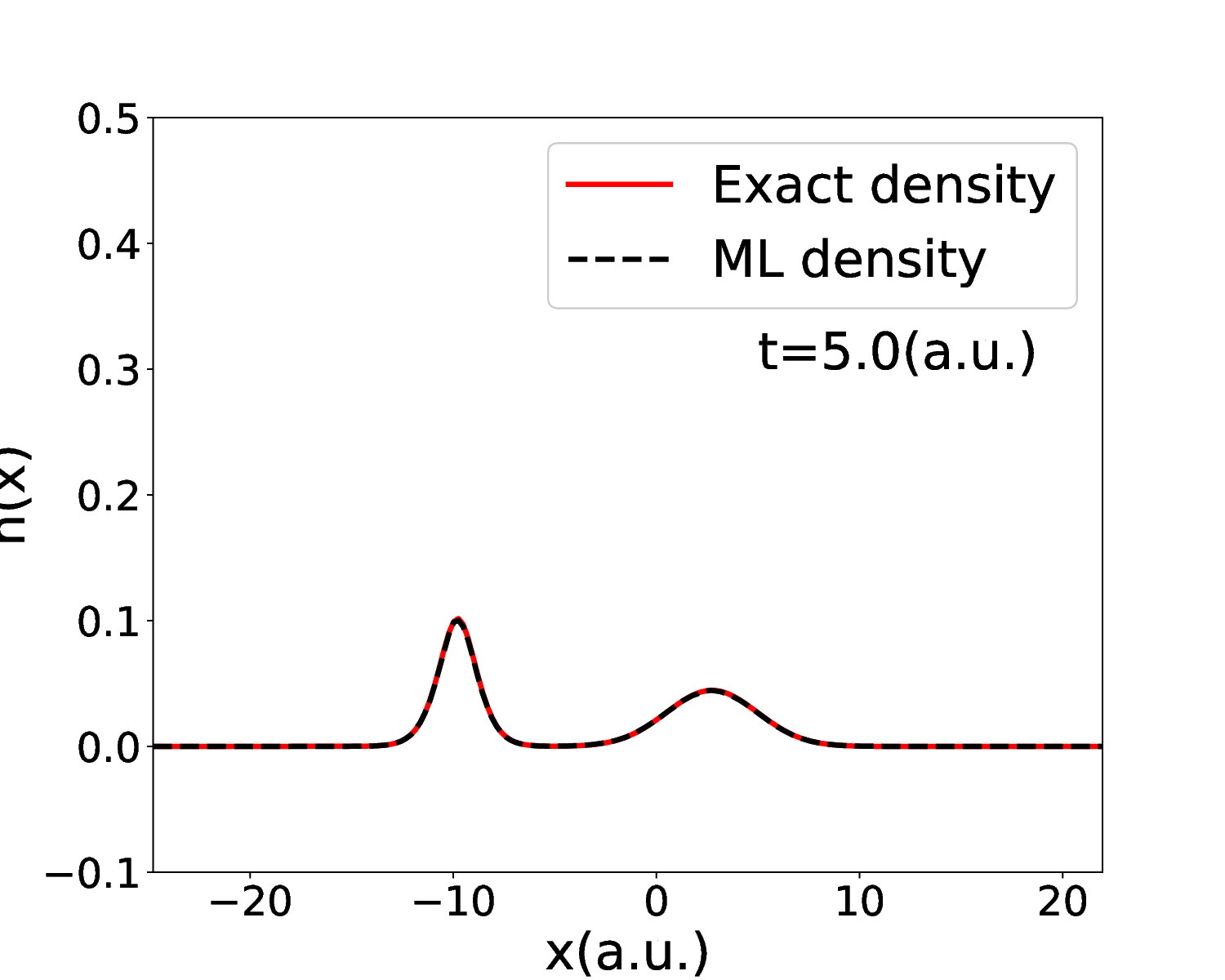}
\end{subfigure}
\begin{subfigure}[b]{0.32\textwidth}
    \centering
    \includegraphics[width=\textwidth]{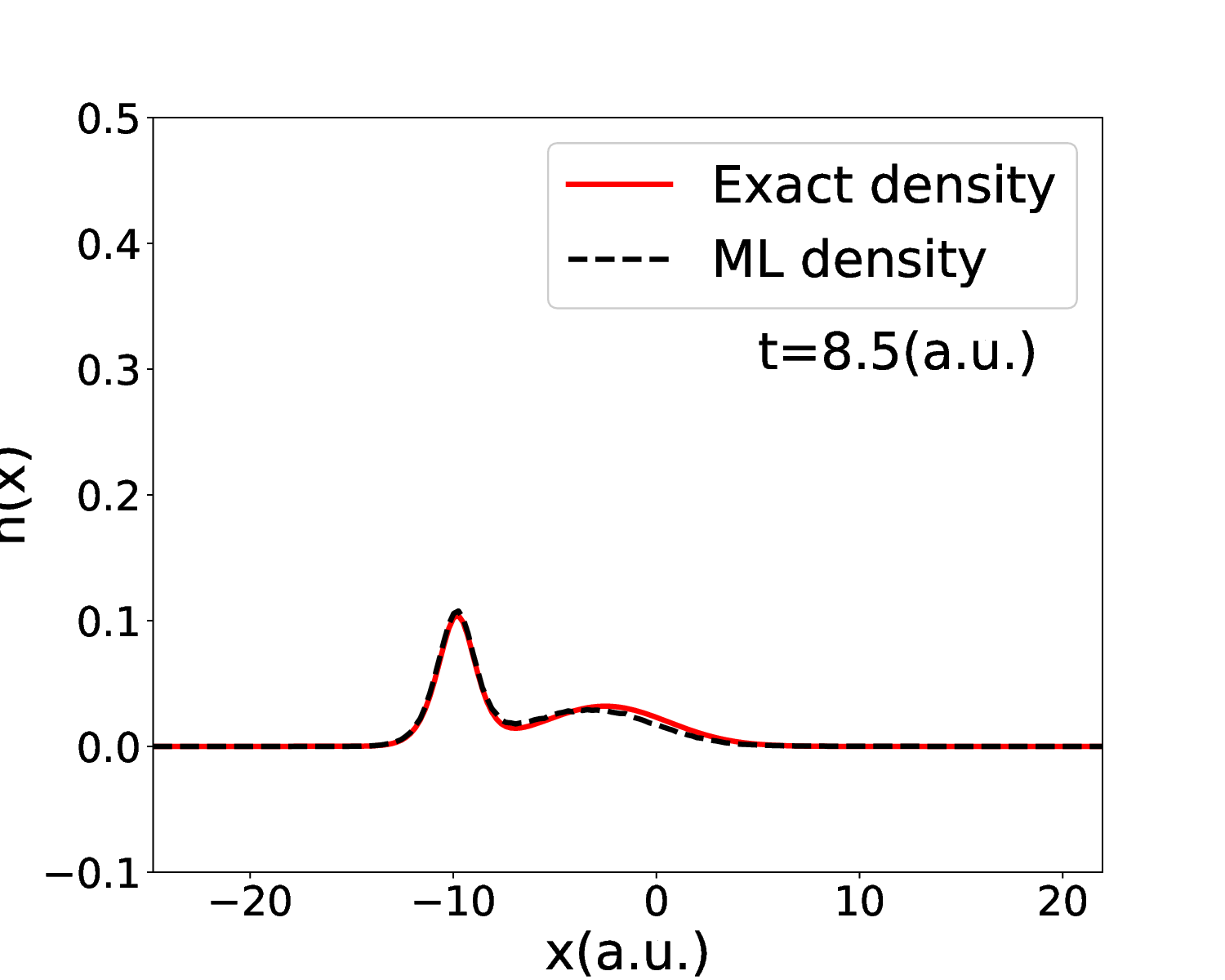}
\end{subfigure}
\begin{subfigure}[b]{0.32\textwidth}
    \centering
    \includegraphics[width=\textwidth]{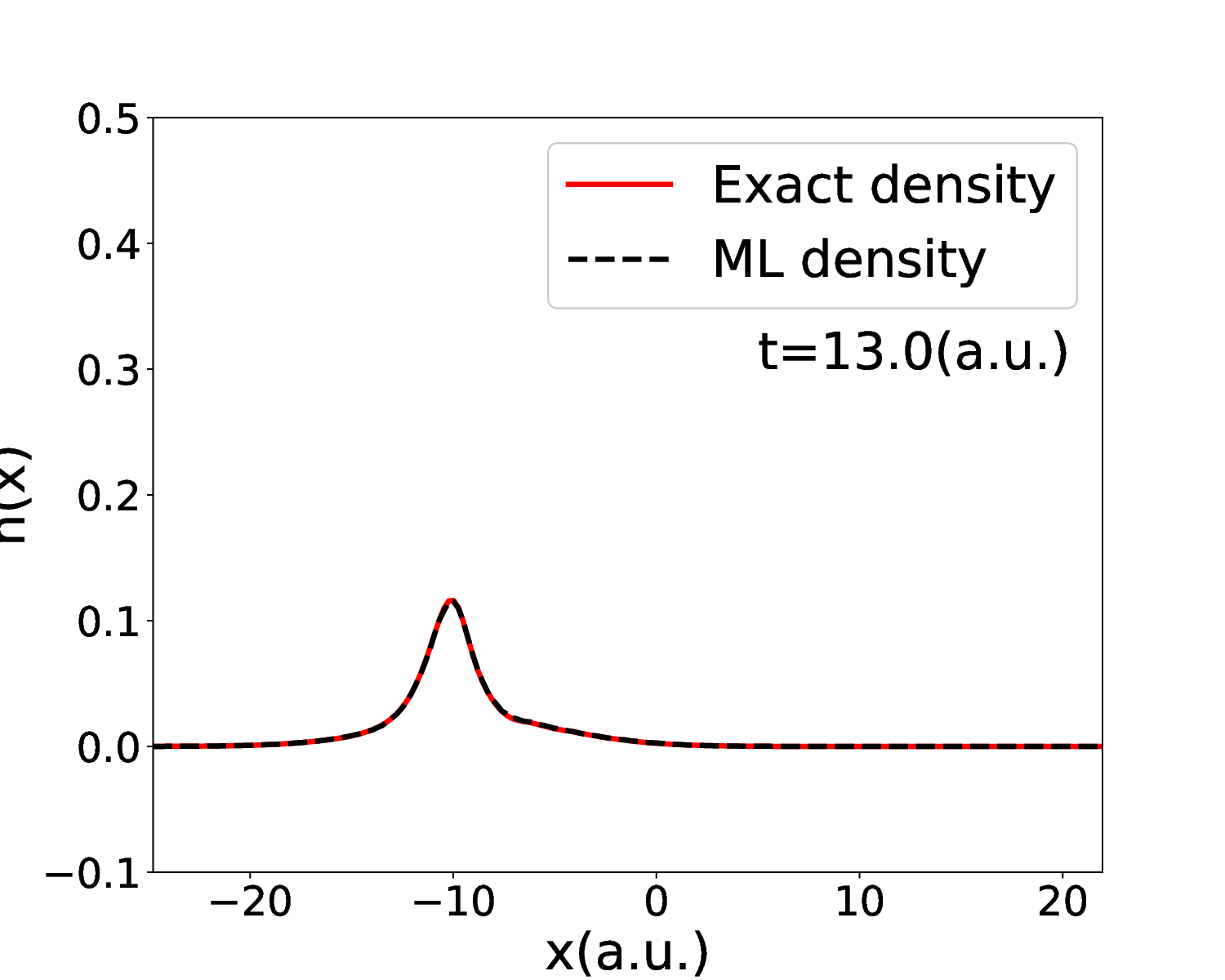}
\end{subfigure}
    \caption{The machine learned  density and the exact density at different timestamps in the two-electron test with a calibration every 500 steps. The black dashed line is the machine-learned density, the red solid line is the exact density. The three panels correspond to the results of timestamp $t = 5.0 \,(a.u.)$ (left), $t = 8.5 \,(a.u.)$ (middle), and $t = 13.0 \,(a.u.)$ (right).}
    \label{fig:2e_evo}
\end{figure}

\begin{figure}[!htbp]
\centering
\begin{subfigure}[b]{0.32\textwidth}
    \centering
    \includegraphics[width=\textwidth]{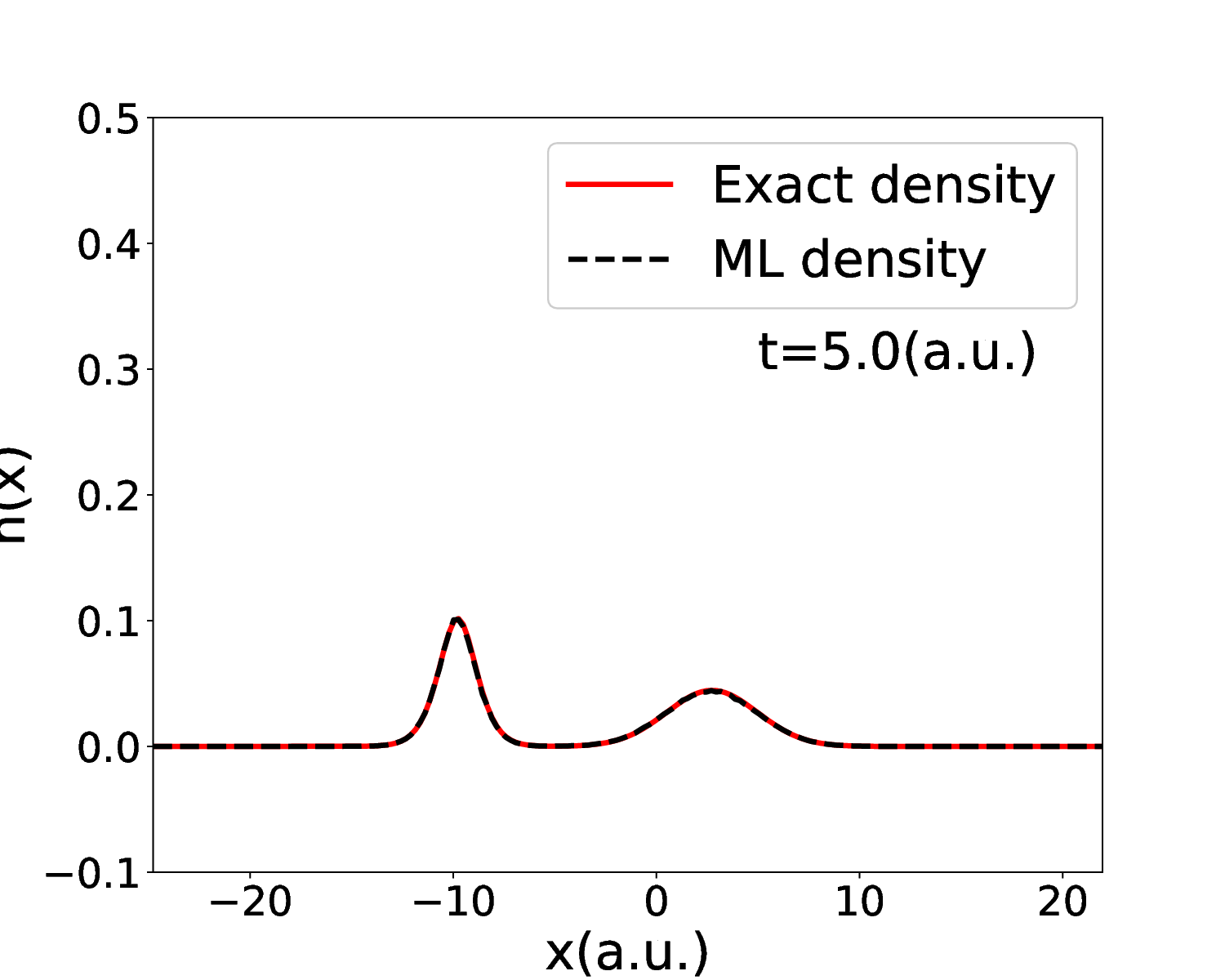}
\end{subfigure}
\begin{subfigure}[b]{0.32\textwidth}
    \centering
    \includegraphics[width=\textwidth]{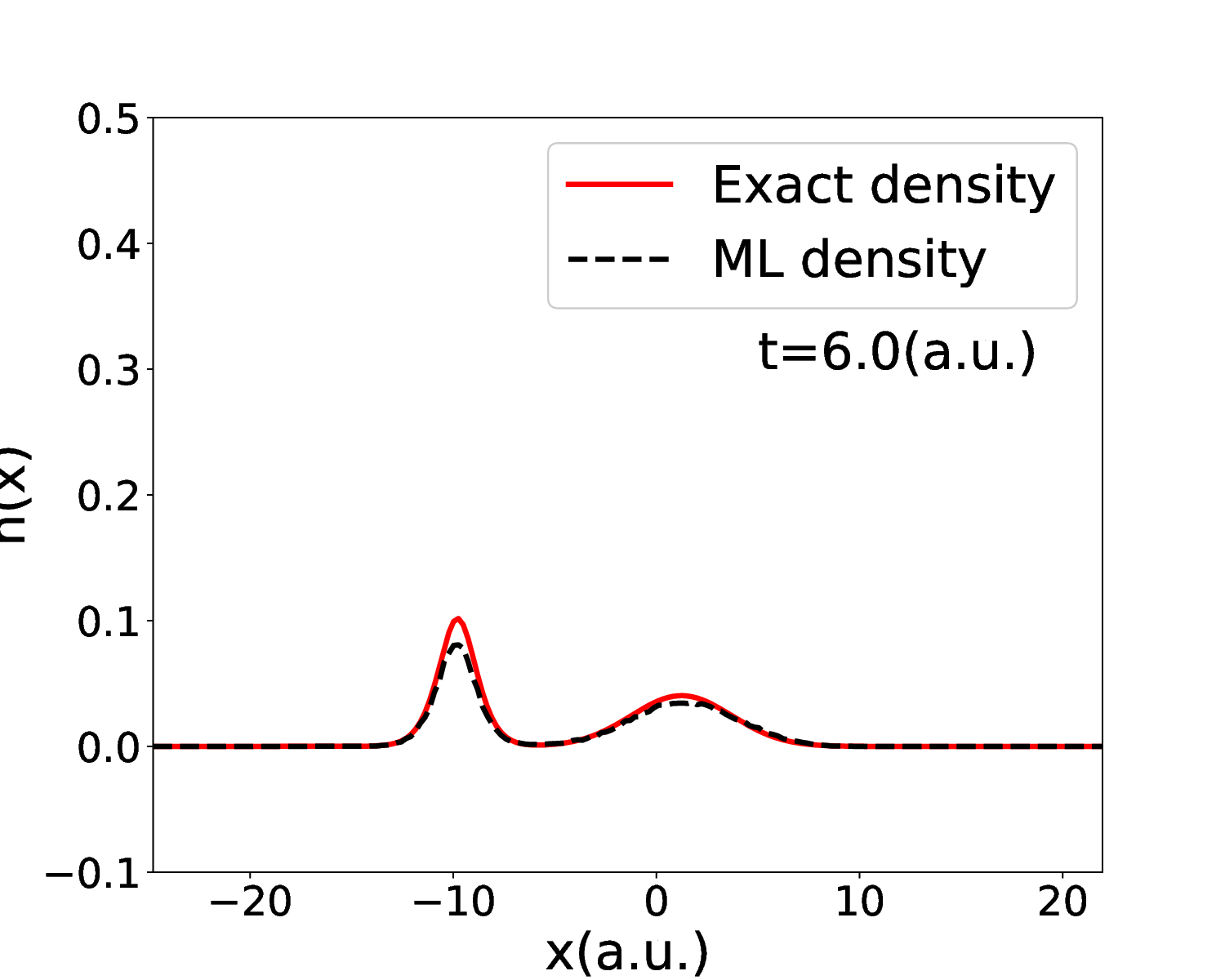}
\end{subfigure}
\begin{subfigure}[b]{0.32\textwidth}
    \centering
    \includegraphics[width=\textwidth]{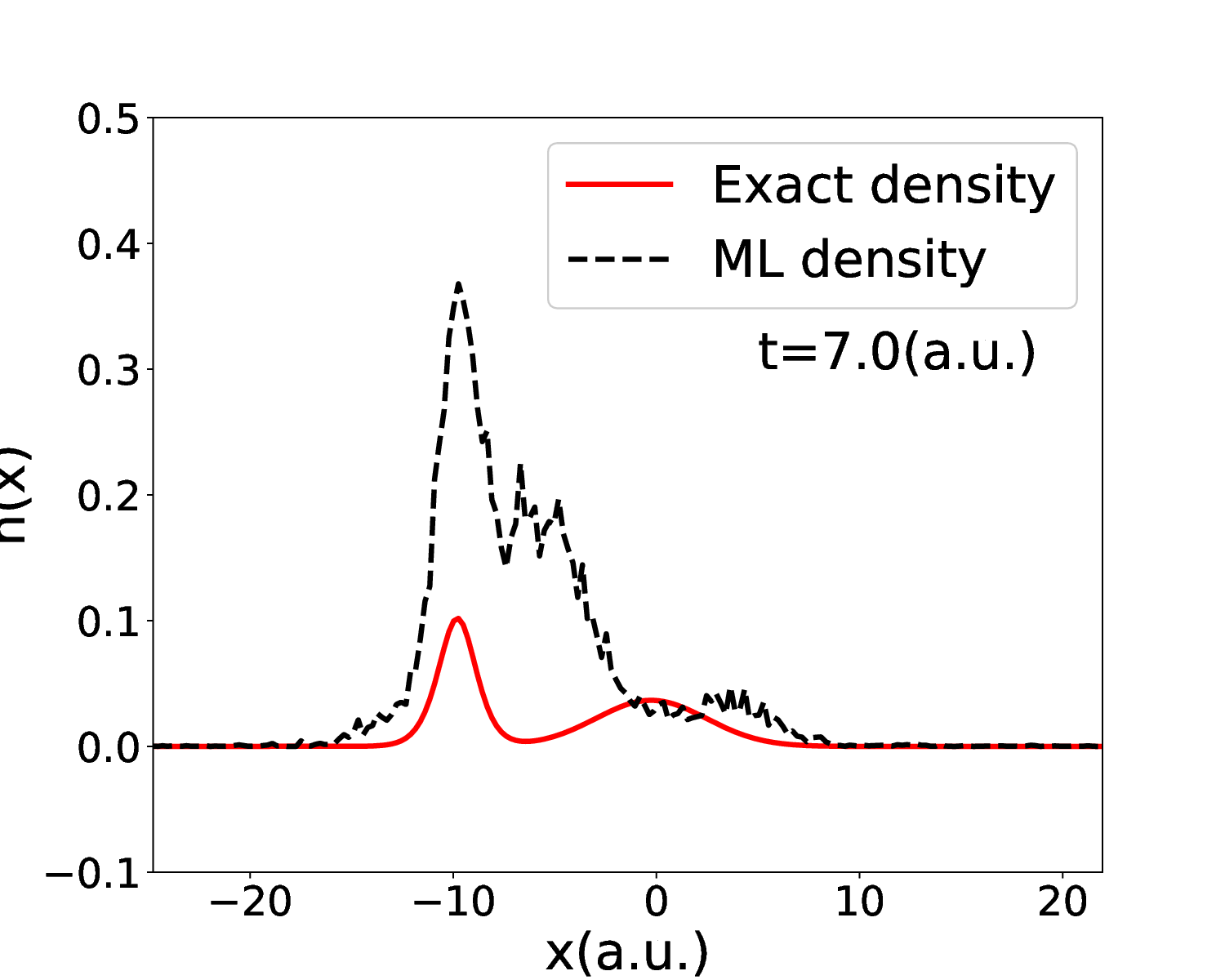}
\end{subfigure}
    \caption{The machine learned  density and the exact density at different timestamps in the two-electron test without calibrations. The black dashed line is the machine-learned density, the red solid line is the exact density. The three panels correspond to the results of timestamp $t = 5.0 \,(a.u.)$ (left), $t = 6.0 \,(a.u.)$ (middle), and $t = 7.0 \,(a.u.)$ (right).}
    \label{fig:2e_evo_no_cali}
\end{figure}

We show the time evolution of the density generated from the neural network in Fig.~\ref{fig:2e_evo}. The time evolution of the machine-learned density (black dashed line) captures the feature of the exact evolution (red solid line). As mentioned in the previous section, the error accumulates over steps, so we calibrate the initial condition every 500 timestamps. We notice that the machine-learned density overlaps well with the exact density at $t = 5.0\, (a.u.)$ and $t = 13.0\, (a.u.)$, but shows a small deviation at $t = 8.5 \,(a.u.)$ when the collision between the two wave packets starts to happen. The choice of calibration every 500 timestamps is to ensure the difference between the machine-learned density and the exact density is small even when the collision happens, which can be increased to a larger number (e.g., 3000) in the absence of collision. As a comparison, we show the results start at $t = 4.0\,(a.u.)$ with no calibration in Fig.~\ref{fig:2e_evo_no_cali}. The error is negligible at $t = 5.0\,(a.u.)$ (1000 steps). It starts to grow at $t=6.0\,(a.u.)$ (2000 steps) and blows up at $t=7.0\,(a.u.)$ (3000 steps). Therefore, calibrations to the initial condition are necessary to prevent the error from accumulating.

\begin{figure}[!htbp]
\centering
\begin{subfigure}[b]{0.45\textwidth}
    \centering
    \includegraphics[width=\textwidth]{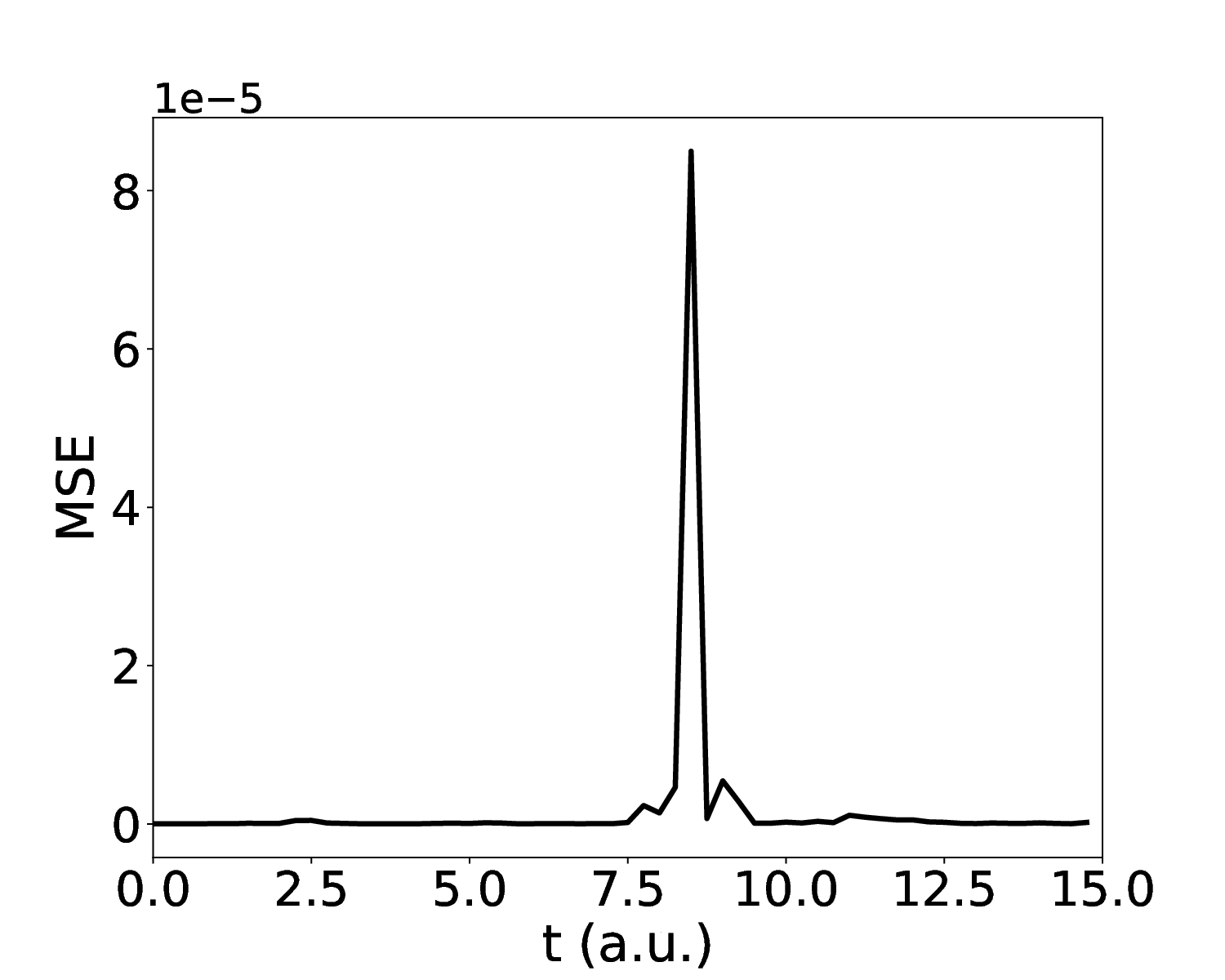}
    \caption{}
    \label{Fig:MSE_2e}
\end{subfigure}
\begin{subfigure}[b]{0.45\textwidth}
    \centering
    \includegraphics[width=0.92\textwidth]{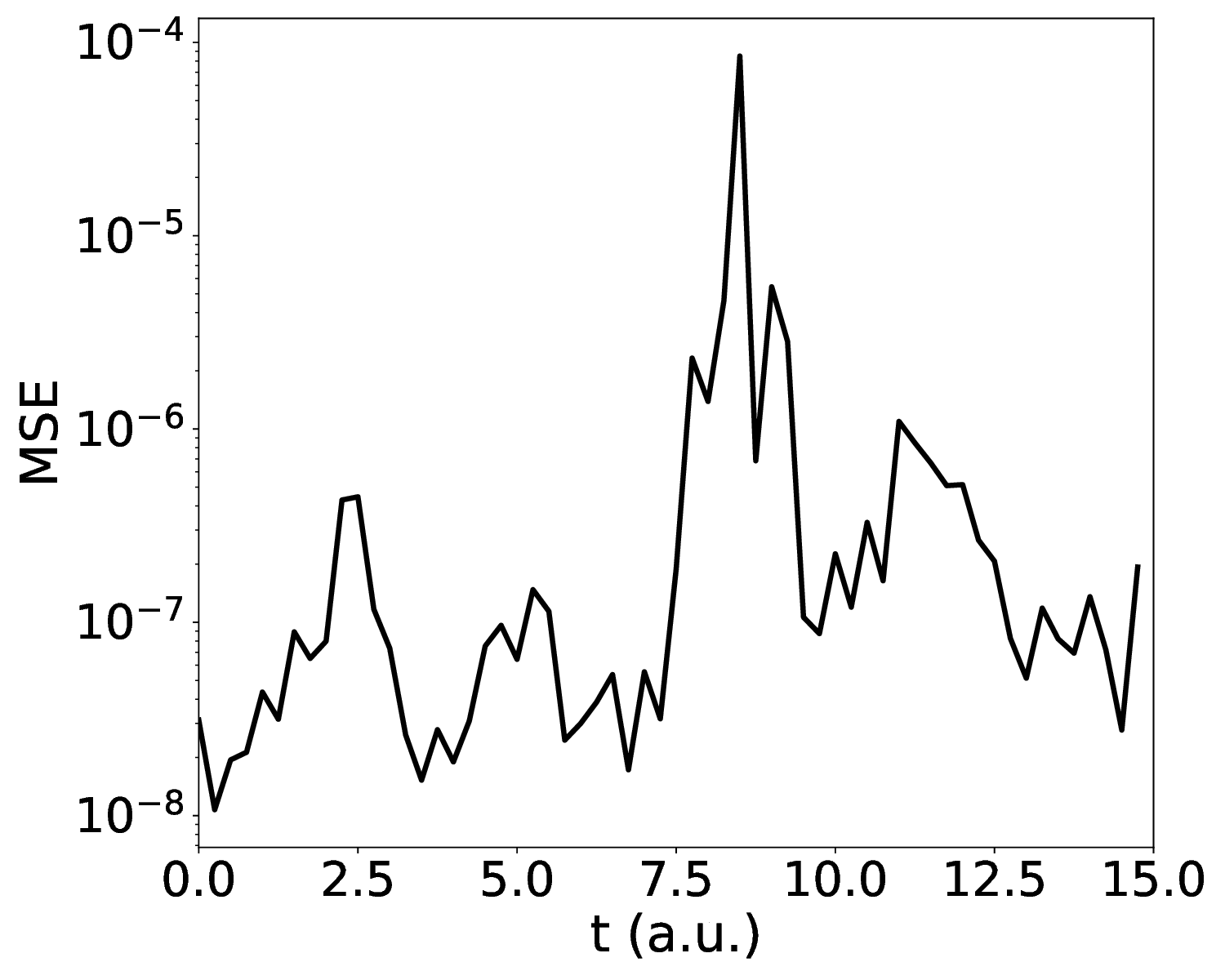}
     \caption{}
     \label{Fig:MSE_2e_logy}
\end{subfigure}
\begin{subfigure}[b]{0.45\textwidth}
    \centering
    \includegraphics[width=\textwidth]{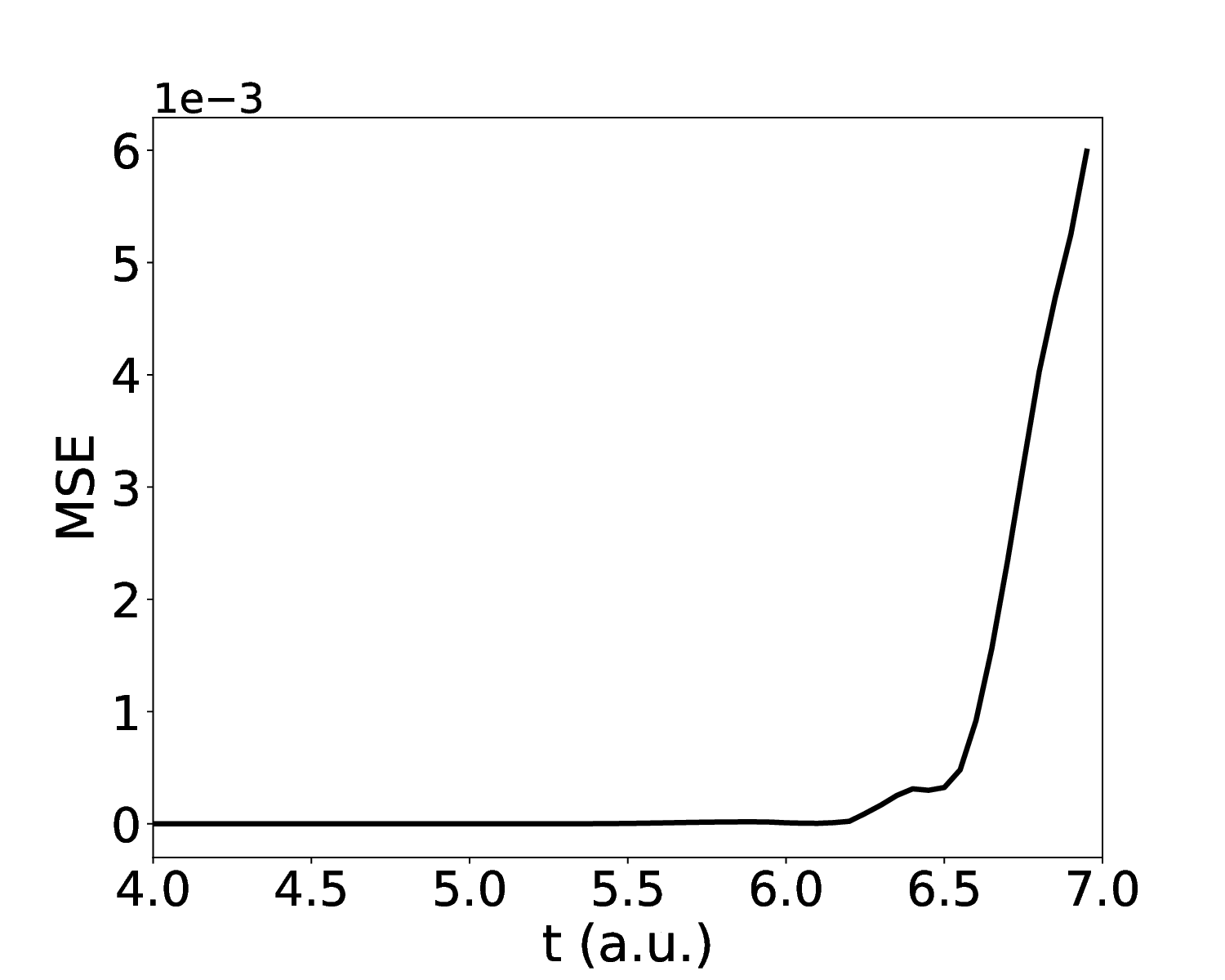}
     \caption{}
     \label{Fig:MSE_2e_nocali}
\end{subfigure}
\begin{subfigure}[b]{0.45\textwidth}
    \centering
    \includegraphics[width=0.92\textwidth]{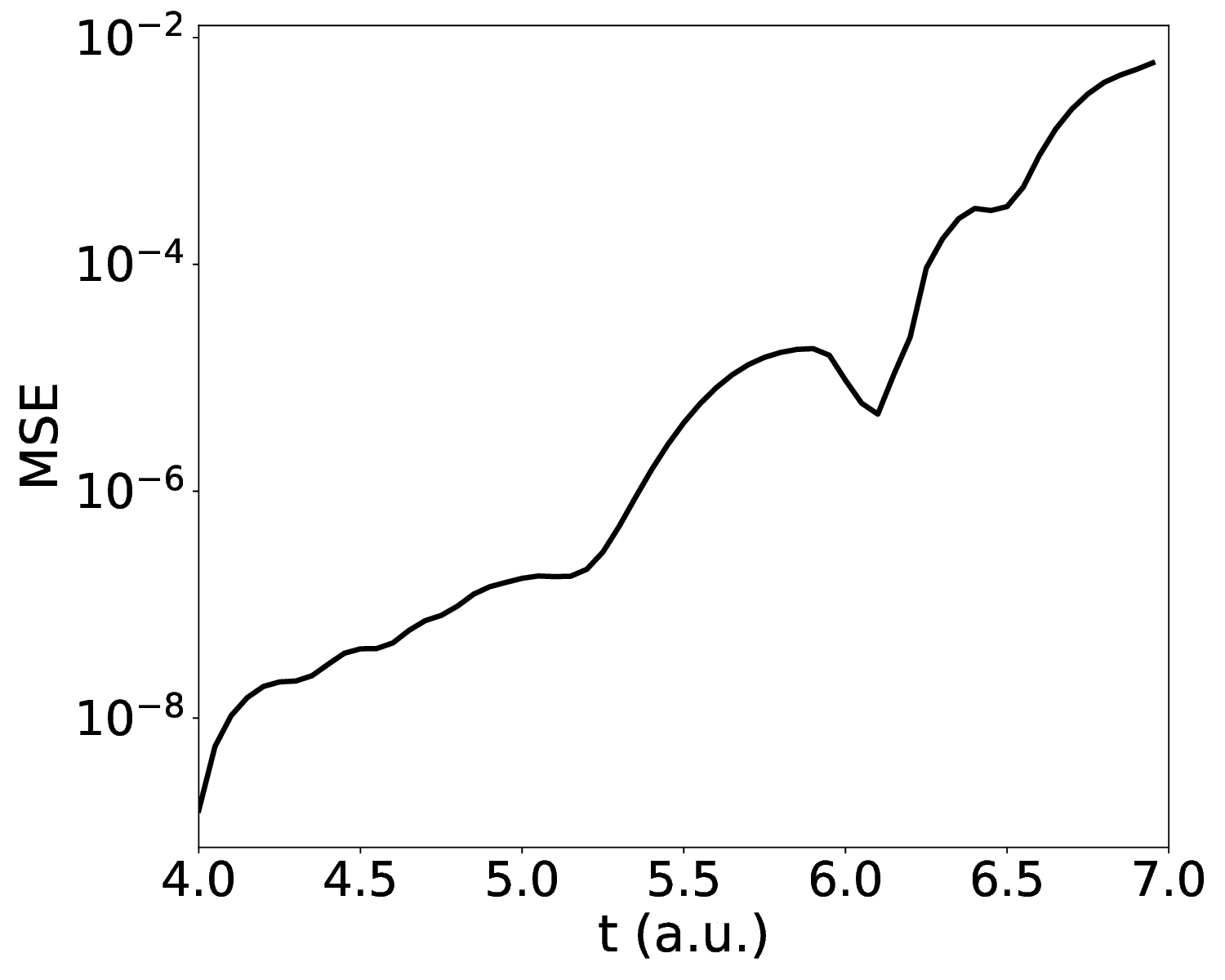}
     \caption{}
     \label{Fig:MSE_2e_nocali_logy}
\end{subfigure}
    \caption{(a) MSE measured at different times in linear scales with calibrations every 500 steps, a spike shows up at $t\approx8.5\,(a.u.)$. (b) MSE measured at different times in logarithmic scales with calibrations every 500 steps. (c) MSE measured at different times in linear scales without calibrations. (d) MSE measured at different times in logarithmic scales without calibrations.}
    \label{fig:2e_err}
\end{figure}

We show the MSE between the machine-learned electron density and the exact electron density at different times in Fig.~\ref{fig:2e_err}. Fig.~\ref{Fig:MSE_2e} and Fig.~\ref{Fig:MSE_2e_logy} are the MSEs in linear and logarithmic scales  with calibration every 500 steps, respectively. Both show a spike at $t\approx 8.5\, (a.u.)$, which is consistent with our abovementioned observations: the error is larger at around $t=8.5 \,(a.u.)$ when the collision between the two wave packets starts to happen. Fig.~\ref{Fig:MSE_2e_nocali} and Fig.~\ref{Fig:MSE_2e_nocali_logy} are MSEs in linear and logarithmic scales without calibrations. We observe that the MSE changes dramatically after $t\approx6.5\,(a.u.)$ in Fig.~\ref{Fig:MSE_2e_nocali}. The relationship between MSE and time becomes evident in the logarithmic scale plot. As shown in Fig.~\ref{Fig:MSE_2e_nocali_logy}, the MSE grows exponentially with time.

\begin{figure}[!htbp]
\centering
\begin{subfigure}[b]{0.32\textwidth}
    \centering
    \includegraphics[width=\textwidth]{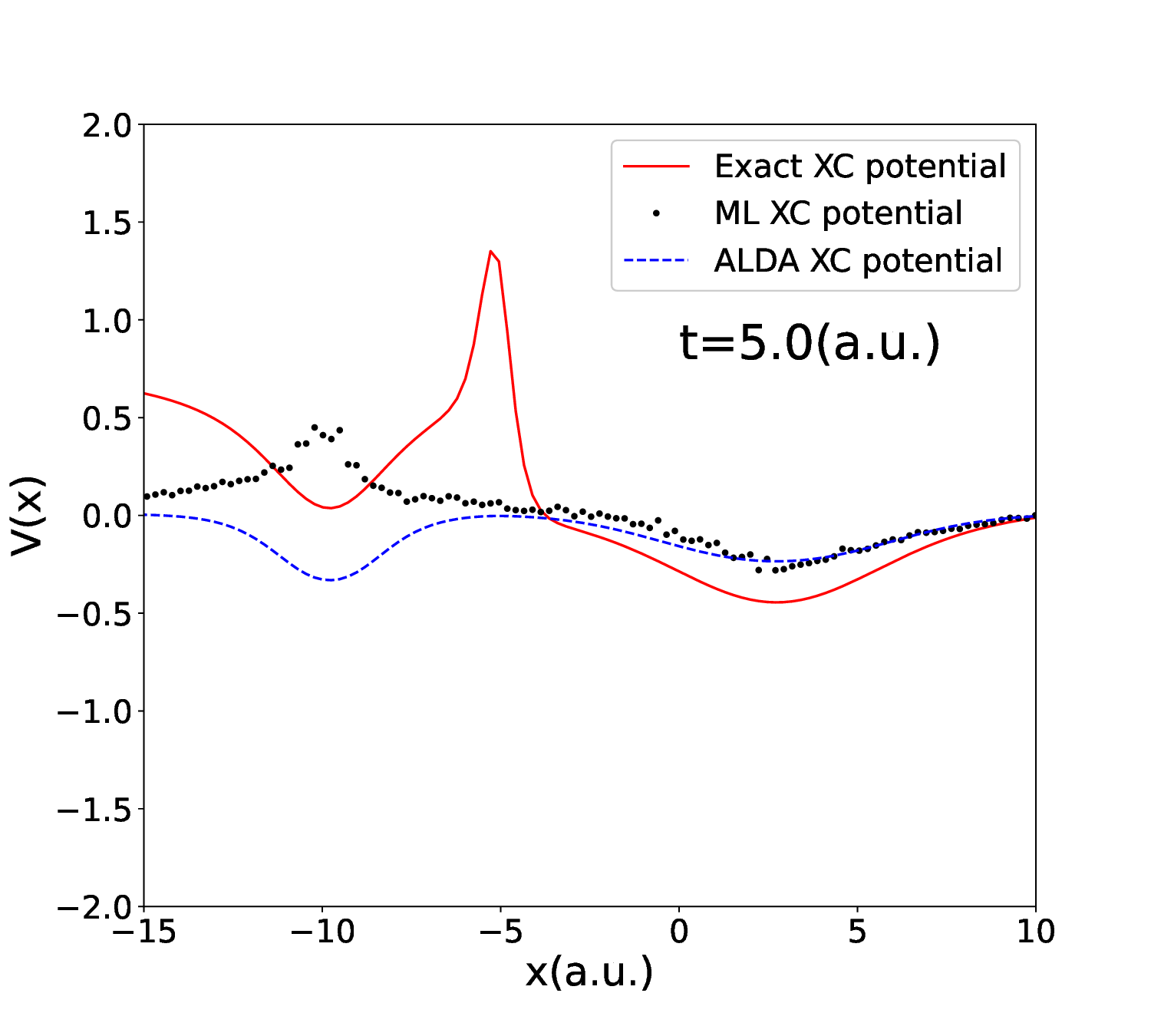}
    \caption{}
    \label{fig:2eks_a}
\end{subfigure}
\begin{subfigure}[b]{0.32\textwidth}
    \centering
    \includegraphics[width=\textwidth]{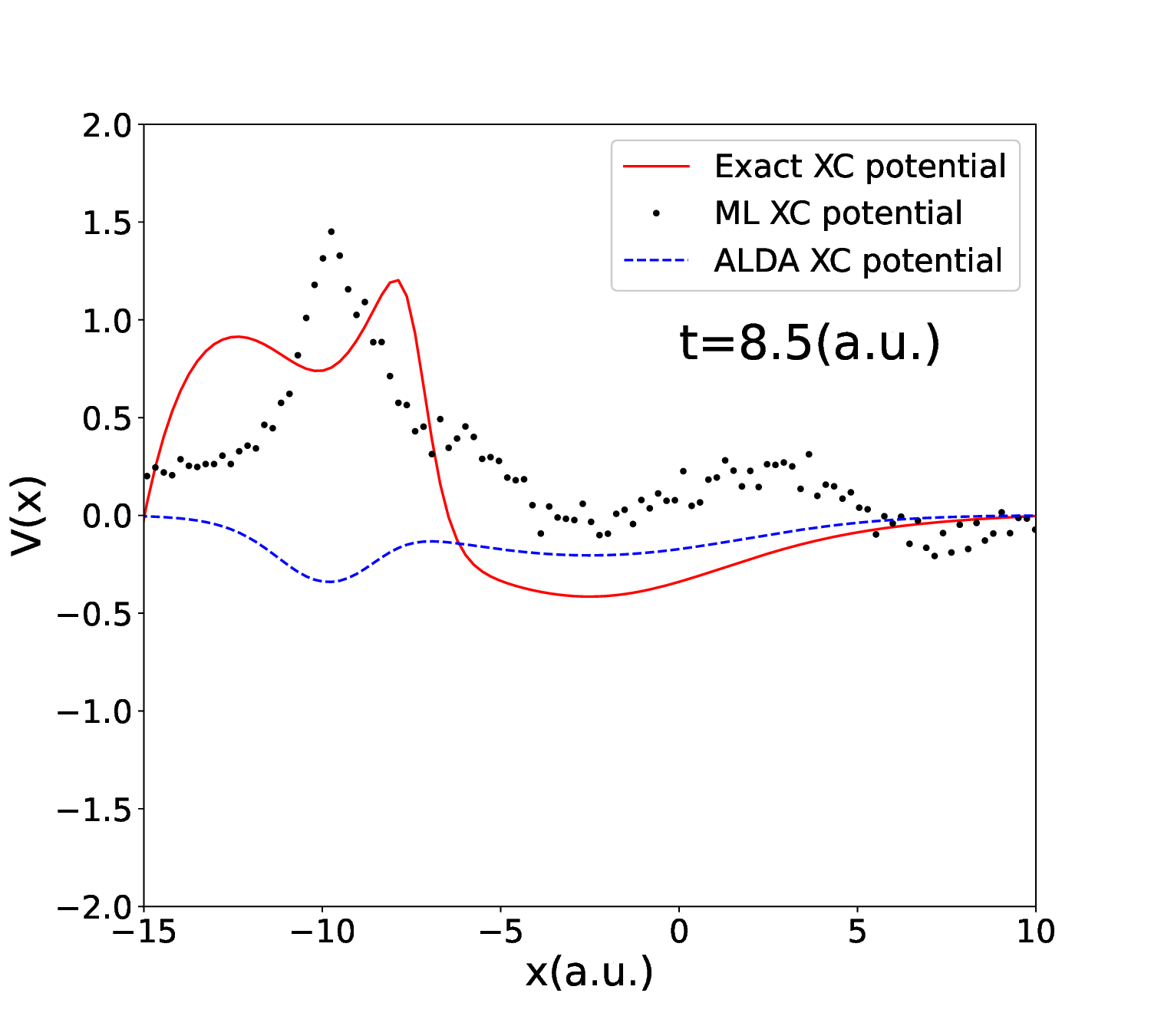}
    \caption{}
    \label{fig:2eks_b}
\end{subfigure}
\begin{subfigure}[b]{0.32\textwidth}
    \centering
    \includegraphics[width=\textwidth]{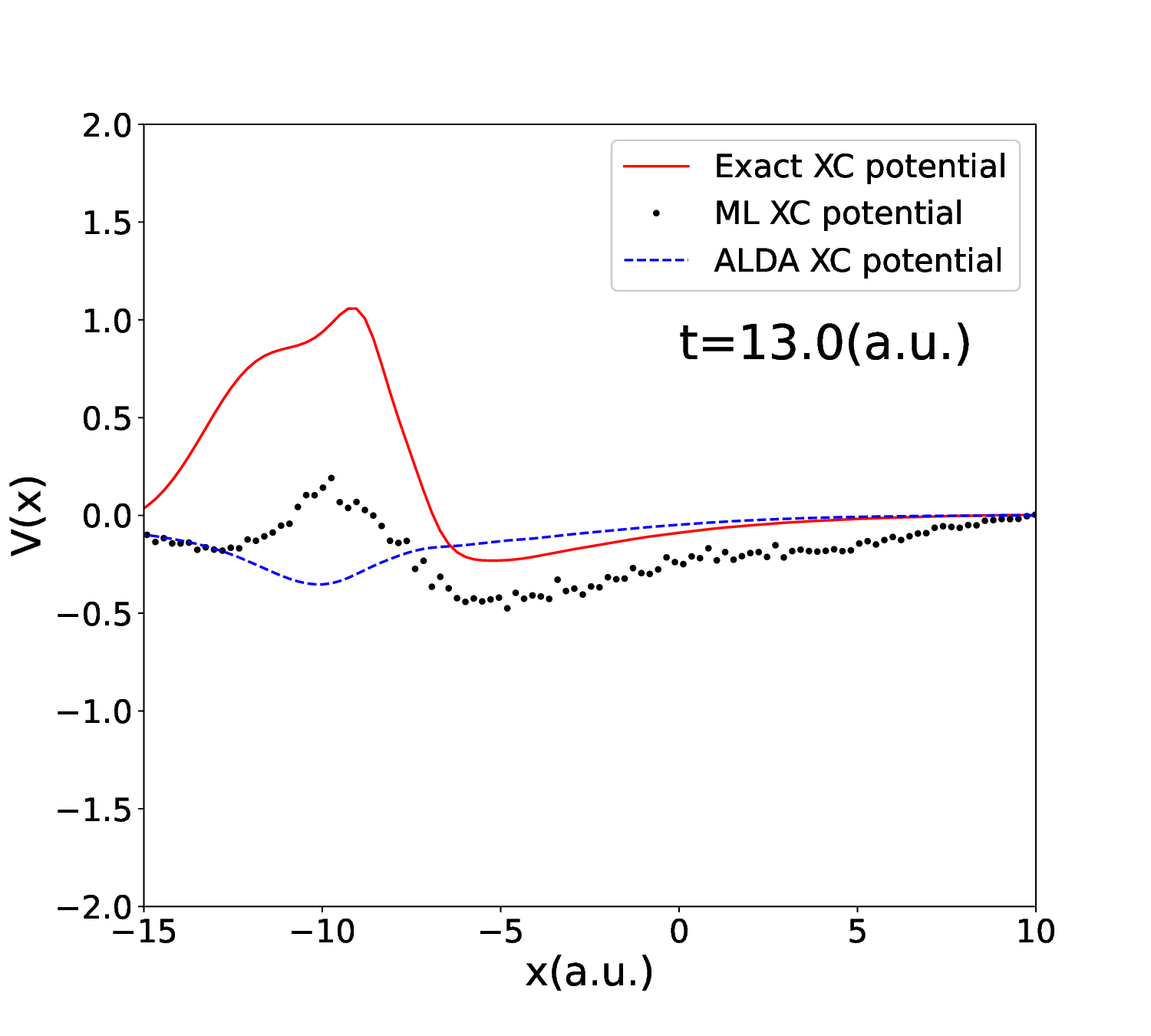}
    \caption{}
    \label{fig:2eks_c}
\end{subfigure}
\begin{subfigure}[b]{0.32\textwidth}
    \centering
    \includegraphics[width=\textwidth]{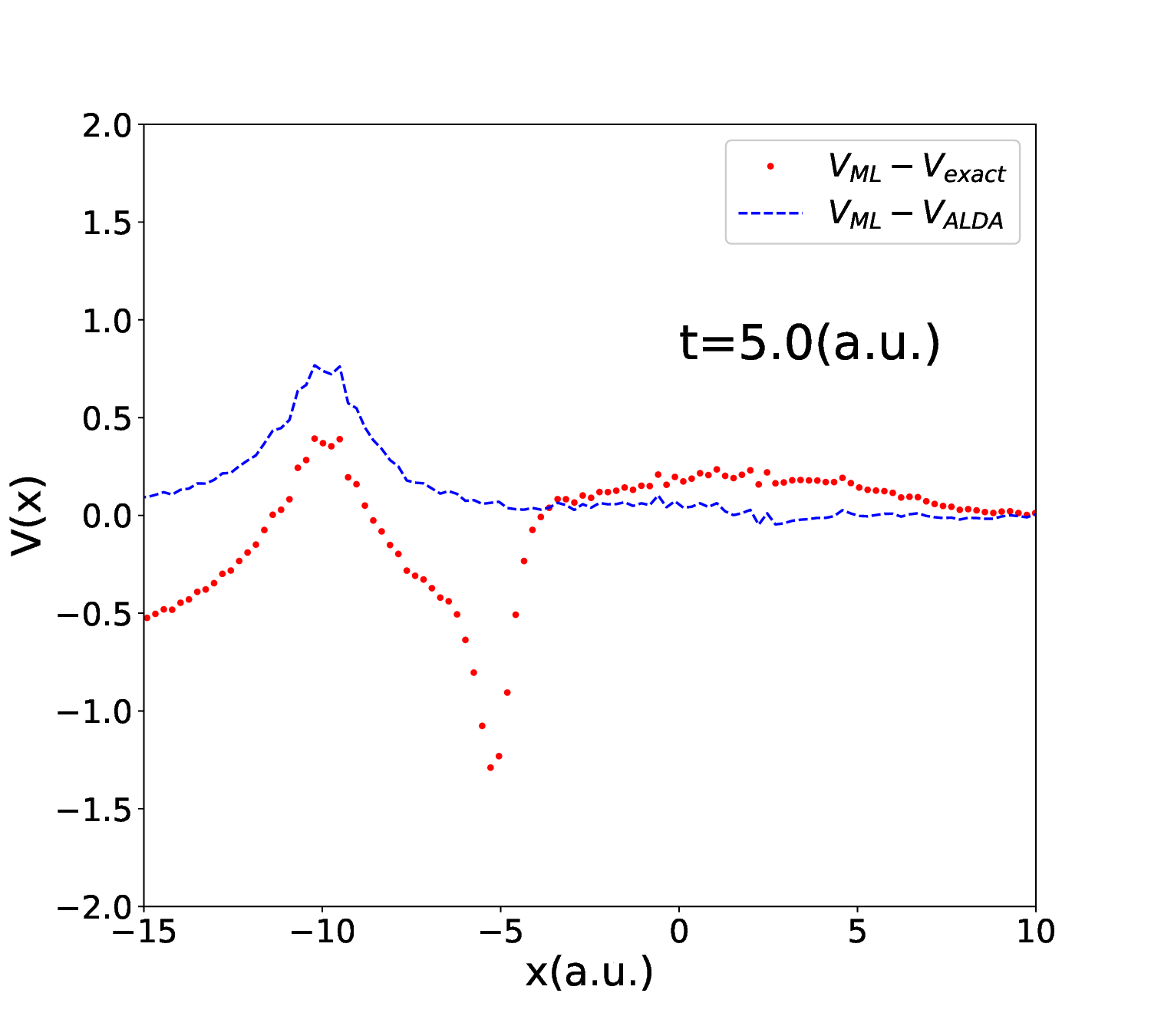}
    \caption{}
    \label{fig:2eks_d}
\end{subfigure}
\begin{subfigure}[b]{0.32\textwidth}
    \centering
    \includegraphics[width=\textwidth]{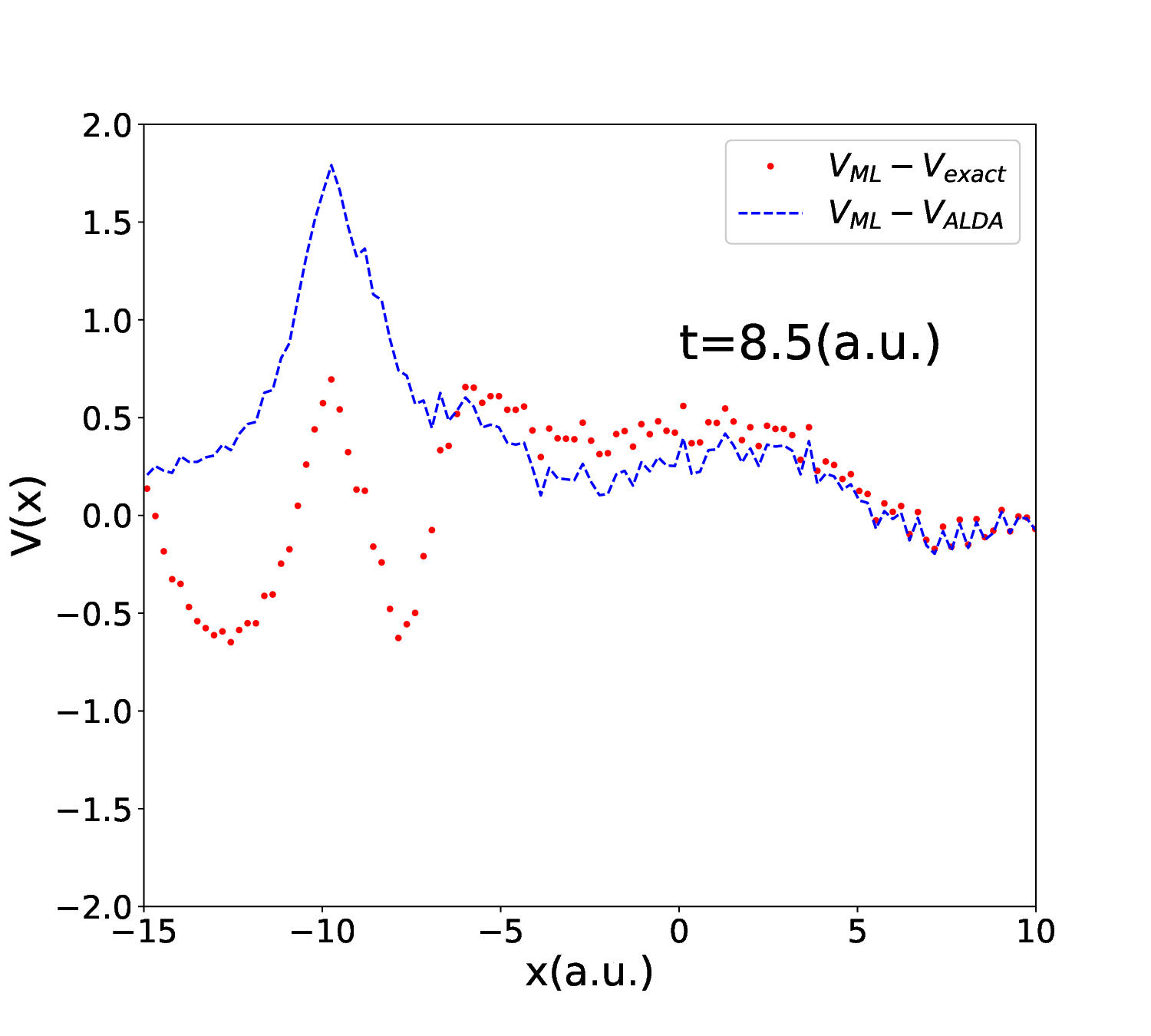}
    \caption{}
    \label{fig:2eks_e}
\end{subfigure}
\begin{subfigure}[b]{0.32\textwidth}
    \centering
    \includegraphics[width=\textwidth]{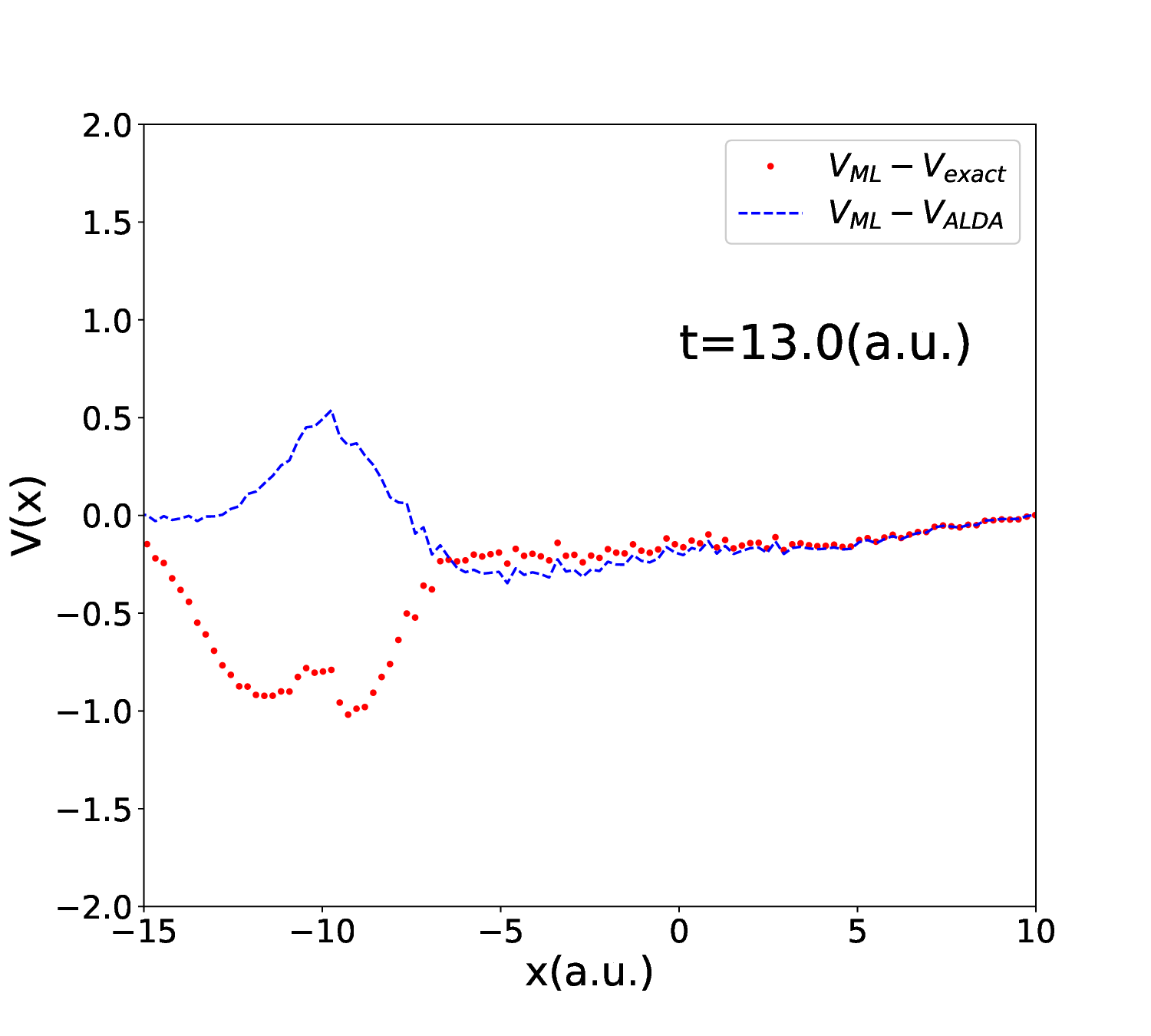}
    \caption{}
    \label{fig:2eks_f}
\end{subfigure}
    \caption{(a) - (c): The machine-learned xc potential, the exact xc potential, and the ALDA xc potential at different timestamps in the two-electron test. The black dashed line is the machine-learned density, the red solid line is the exact density, and the blue dashed line is the ALDA potential. 
    (d) - (f): The red dotted line is the difference between the machine-learned xc potential and the exact xc potential. The blue dashed line is the difference between the machine-learned xc potential and the ALDA xc potential.
    The three columns correspond to the result at $t = 5.0 \,(a.u.)$ (left), $t = 8.5 \,(a.u.)$ (middle), and $t = 13.0 \,(a.u.)$ (right).}
    \label{fig:2eks_potential}
\end{figure}

Although the neural network reproduces the correct dynamics of the system, we need to mention that the machine-learned xc potential is not the same as the actual one. We show the comparison in Fig.~\ref{fig:2eks_potential}. One note here is that we don't compare the xc potential directly. Instead, we compare the exchange-correlation part of the xc potential.
 The machine-learned xc potential (black dots) does not capture the features of the actual xc (red solid line) potential. Similar phenomena were observed in Ref.~\cite{Bhat2021Dec}, where the dynamics of the same system were reproduced correctly by adjoint learning, but they mentioned that the machine-learned xc potential did not correspond to the actual one. The adiabatic local density approximation (ALDA) xc potential (blue dash line) is also shown for comparison. Due to the lack of memory effect, the ALDA xc potential does not capture the exact dynamics of the two-electron system.
 
 In our machine learning method, we attempted to learn the correct dynamics under adiabatic approximation. So, the machine-learned xc potential is expected to be different from both the exact xc potential and the ALDA xc potential but to retain the features of both potentials. The results shown in Fig.~\ref{fig:2eks_potential} can show some evidence. In Fig.~\ref{fig:2eks_a} and Fig.~\ref{fig:2eks_b}, the machine-learned xc potential peaks at $x=-10$, which is in alignment with the ALDA xc potential. The trend of machine-learned xc potential is similar to the exact xc potential in the region of $x > 0$. The difference between the exact xc potential and the machine-learned potential, as well as the difference between the machine-learned and the ALDA potential, are shown in Figs.~\ref{fig:2eks_d},\ref{fig:2eks_e} and \ref{fig:2eks_f}.

We analyzed the entire procedure, and the most important reason for the discrepancy between the machine-learned xc potential and the actual one is the memory effect. It has been proven that the xc potential at time $t$ in this two-electron system depends on all densities at $t^\prime < t$ and the choice of the initial Kohn-Sham orbitals\cite{Suzuki2017Dec, Suzuki2020May, Lacombe2018Jun, Maitra2002Jun}. In our work, we assume adiabatic approximation, so the dynamics and the xc potential of the system cannot be reproduced correctly at the same time.
\section{Conclusions}\label{conclusion}
In this article, we have described a new machine-learning method to learn the dynamics and the Kohn-Sham potential (or the hardest part -- xc potential) of the Kohn-Sham system. We have demonstrated the method with two one-dimensional examples: a harmonic oscillator model and a two-electron soft Coulomb model. In both examples, the exact dynamics of the systems could be well reproduced from the machine learning method. The machine-learned  potential in the harmonic oscillator test captures the general feature of the actual quadratic form potential, but it shows a discrepancy from the actual one in the two-electron test. We have analyzed the possible reasons, and the memory effect is the major source of error. The memory effect requires considering the densities of previous timestamps. To overcome this difficulty, we believe the neural networks capable of handling time series (e.g., RNN, LSTM) are promising\cite{Sherstinsky2018Aug, Hochreiter1997Nov}.
\section{Acknowledgements}
JY and JDW were supported by the U.S. Department of Energy, Office
of Science, and Office of Advanced Scientific Computing Research, under the Quantum
Computing Application Teams program (Award 1979657).  JDW was also supported by the NSF (Grant 1820747) and additional funding from the DOE (Award A053685). 
   
\appendix
\section{Schr\"{o}dinger equation and Hamilton's equations}\label{pf:se_he}
In this part, we show the relation between the Schr\"{o}dinger equation and Hamilton's equations. For convenience, we want to write Hamilton's equation into a compact form in terms of complex variables. In classical mechanics, given a Hamiltonian $H$, the dynamics of the system can be determined by Hamilton's equation:
\begin{eqnarray}
    &\frac{\partial H}{\partial p} = \dot{q} \\
    &\frac{\partial H}{\partial q} = -\dot{p},
\end{eqnarray} where $q$ and $p$ are the canonical coordinates in phase space.
By rewriting $z = \frac{1}{\sqrt{2}}(q + ip)$ and $z^* = \frac{1}{\sqrt{2}}(q - ip)$, we can transform the derivatives into the following form:
\begin{eqnarray}
    \frac{\partial}{\partial z} = \frac{\partial}{\partial q}\frac{\partial q}{\partial z} +  \frac{\partial}{\partial p}\frac{\partial p}{\partial z} = \frac{1}{\sqrt{2}}(\frac{\partial}{\partial q} - i\frac{\partial}{\partial p})\\
    \frac{\partial}{\partial z^*} = \frac{\partial}{\partial q}\frac{\partial q}{\partial z^*} +  \frac{\partial}{\partial p}\frac{\partial p}{\partial z^*} = \frac{1}{\sqrt{2}}(\frac{\partial}{\partial q} + i\frac{\partial}{\partial p}).
\end{eqnarray}
Thus we can write Hamilton's equations in a tighter form,
\begin{equation}
    i\dot{z} = \frac{\partial H}{\partial z^*}.
\end{equation}
Now let's look at the Schr\"{o}dinger's equation:
\begin{equation}
i\dt\psit = \hat{H}\psit.
\end{equation}
Given the basis states $\ket{k}, k=0, 1,\ldots$, the wave function can be expanded as
\begin{equation}
    \psit = \sum_k c_k(t) \ket{k}.
\end{equation}

Thus, we obtain the differential equation of the coefficients vector, $\bi{c}(t) = [c_1(t), c_2(t), \cdots, c_n(t)]^T$
\begin{equation}
    i\dot{\bi{c}} = \bi{Hc},
\end{equation} where $\bi{H}$ is the matrix representation of the Hamiltonian $\hat{H}$ with its entries being $H_{ij} = \braket{i|\hat{H}|j}$.

By defining $H(\bi{c})  = \braket{\Psi(t)|\hat{H}|\Psi(t)} = \bi{c^\dagger Hc}$\cite{Strocchi1966Jan}, the Schr\"{o}dinger's equation can be transformed as,
\begin{equation}
    i\dot{\bi{c}} = \frac{\partial H(\bi{c})}{\partial \bi{c}^\dagger},
\end{equation} which is consistent with the classical Hamilton's equation.
\section{Kohn-Sham equations and Hamilton's equation}\label{pf:ks_he}
We prove Hamilton's equations in Kohn-Sham system in this part. The Kohn-Sham equations are given by,
\begin{eqnarray}
    i\frac{\partial }{\partial t} \ket{\phi_m(t)} = \hat{H}_{KS}[n(t)] \ket{\phi_m(t)}, m = 1, 2, \ldots,
\end{eqnarray} where $N$ is the number of electrons, and $n(t) = \sum_{m=1}^N |\phi_m(t)|^2$.

In section~\ref{section:KS_HE}, we have shown the Kohn-Sham equation can be written as a differential equation in terms of the vector coefficients $\bi{c}_m$, given the basis functions $\{s_i(\bi{r}) = \braket{\bi{r}|i}\}$:
\begin{equation}
    i\dot{\bi{c}}_m = \bi{H}_{KS}[\bi{c}]\bi{c}_m,
\end{equation} where the matrix element at $i$-th row $j$-th column is given by:
\begin{equation}
    (\bi{H}_{KS}[\bi{c}])_{ij} = \braket{i|\hat{T}_s[n(t)] + \hat{V}_H[n(t)] + \hat{V}_{ext}[n(t)] + \hat{V}_{xc}[n(t)]|j}. 
\end{equation}

To bridge Kohn-Sham equations and Hamilton's equation, we need to find an energy functional $H_{KS}[\bi{c}]$, such that $\frac{\partial H_{KS}[\bi{c}]}{\partial \bi{c}^\dagger_m} = \bi{H}_{KS}[\bi{c}] \bi{c}_m$.

Under the adiabatic approximation, the energy functional corresponding to the system is $H_{KS}[n] = T_S[n] + E_{ext}[n] + E_{H}[n] + E_{xc}[n]$, where $n(\bi{c}; \bi{r}) = \sum_{m, i, j} c^*_{im}c_{jm}s_i^*(\bi{r})s_j(\bi{r})$ is the electron density mentioned in the main text.

Therefore, the four terms in the energy functional expansion can be calculated below:
\begin{enumerate}
    \item Kinetic energy:
    \begin{eqnarray}
        &T_s[n] = \sum_m \sum_{i,j} \braket{i|\hat{T}_{s}|j}c_{jm}c^*_{im}\\
        &\frac{\partial T_{s}[n]}{\partial c^*_{im}} = \sum_j\braket{i|\hat{T}_{s}|j}c_{jm}
    \end{eqnarray}
    \item External energy:
    \begin{eqnarray}
    &E_{ext}[n] = \sum_m \sum_{i,j} \braket{i|\hat{V}_{ext}|j}c_{jm}c^*_{im}\\
    &\frac{\partial E_{ext}[n]}{\partial c^*_{im}} = \sum_j\braket{i|\hat{V}_{ext}|j}c_{jm}
    \end{eqnarray}
    \item Hartree energy:
    \begin{eqnarray}
    \nonumber\frac{\partial E_{H}[n]}{\partial c^*_{im}} &= \int\mathrm{d}\bi{r}\int\mathrm{d}\bi{r}^\prime \frac{\delta E_{H}[n]}{\delta n(\bi{c};\bi{r}^\prime)} \delta(\bi{r}-\bi{r}^\prime)\frac{\partial n(\bi{c}; \bi{r})}{\partial c^*_{im}}\\
    &= \sum_{j}\int\mathrm{d}\bi{r} \ v_{H}(\bi{r}) c_{jm}s^*_i(\bi{r})s_j (\bi{r}) = \sum_j\braket{i|\hat{V}_{H}|j}c_{jm}
    \end{eqnarray}
    \item Exchange-correlation energy: 
    \begin{eqnarray}
    \nonumber\frac{\partial E_{xc}[n]}{\partial c^*_{im}} &= \int\mathrm{d}\bi{r}\int\mathrm{d}\bi{r}^\prime \frac{\delta E_{xc}[n]}{\delta n(\bi{c}; \bi{r}^\prime)} \delta(\bi{r}-\bi{r}^\prime)\frac{\partial n(\bi{c}; \bi{r})}{\partial c^*_{im}}\\
    &= \sum_{j}\int\mathrm{d}\bi{r} \ v_{xc}(\bi{r}) c_{jm}s^*_i(\bi{r})s_j (\bi{r}) = \sum_j\braket{i|\hat{V}_{xc}|j}c_{jm}
    \end{eqnarray}
\end{enumerate}

Putting together the three terms, we have
\begin{equation}
\frac{\partial H_{KS}(\bi{c})}{\partial \bi{c^*}_m} = \bi{H}_{KS}[\bi{c}] \bi{c}_m = i \dot{\bi{c}}_m
\end{equation}
This concludes the proof of Eq.~\ref{eq:HE_KS}.

We can take a further step to show Eq.~\ref{eq:v_ks} $\frac{\partial^2 H_{KS}(\bi{c})}{\partial c^*_{mi}\partial c_{mj}} -\braket{i|\hat{T}_s|j} = \braket{i|\hat{V}_{KS}|j}$.
\section{Machine-learned potential scaling with number of eigenstates in the dataset}\label{app: v_scale}

\begin{figure}[!htbp]
\centering

\begin{subfigure}[b]{0.32\textwidth}
    \centering
    \includegraphics[width=\textwidth]{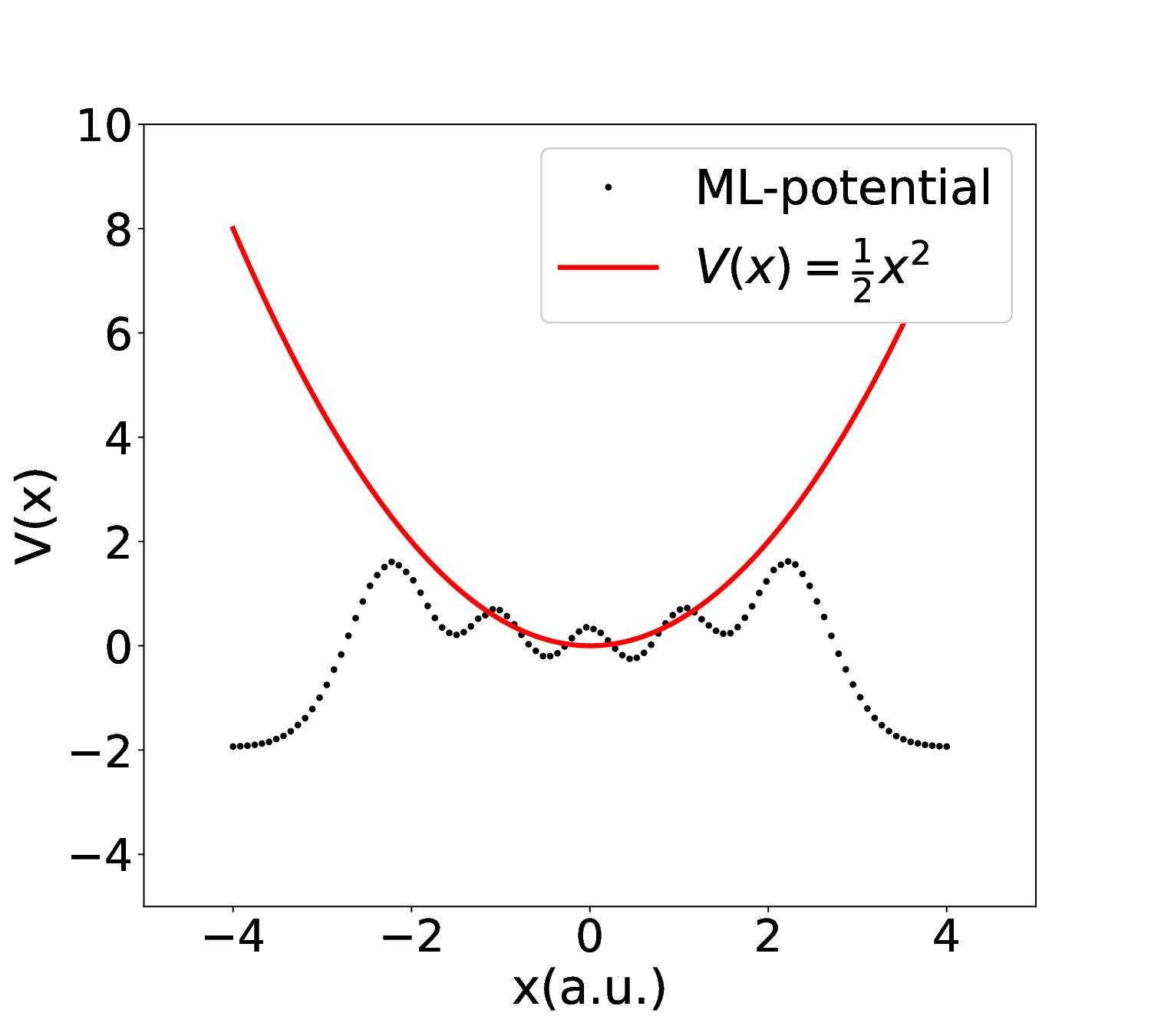}
    \caption{}
\end{subfigure}
\begin{subfigure}[b]{0.32\textwidth}
    \centering
    \includegraphics[width=\textwidth]{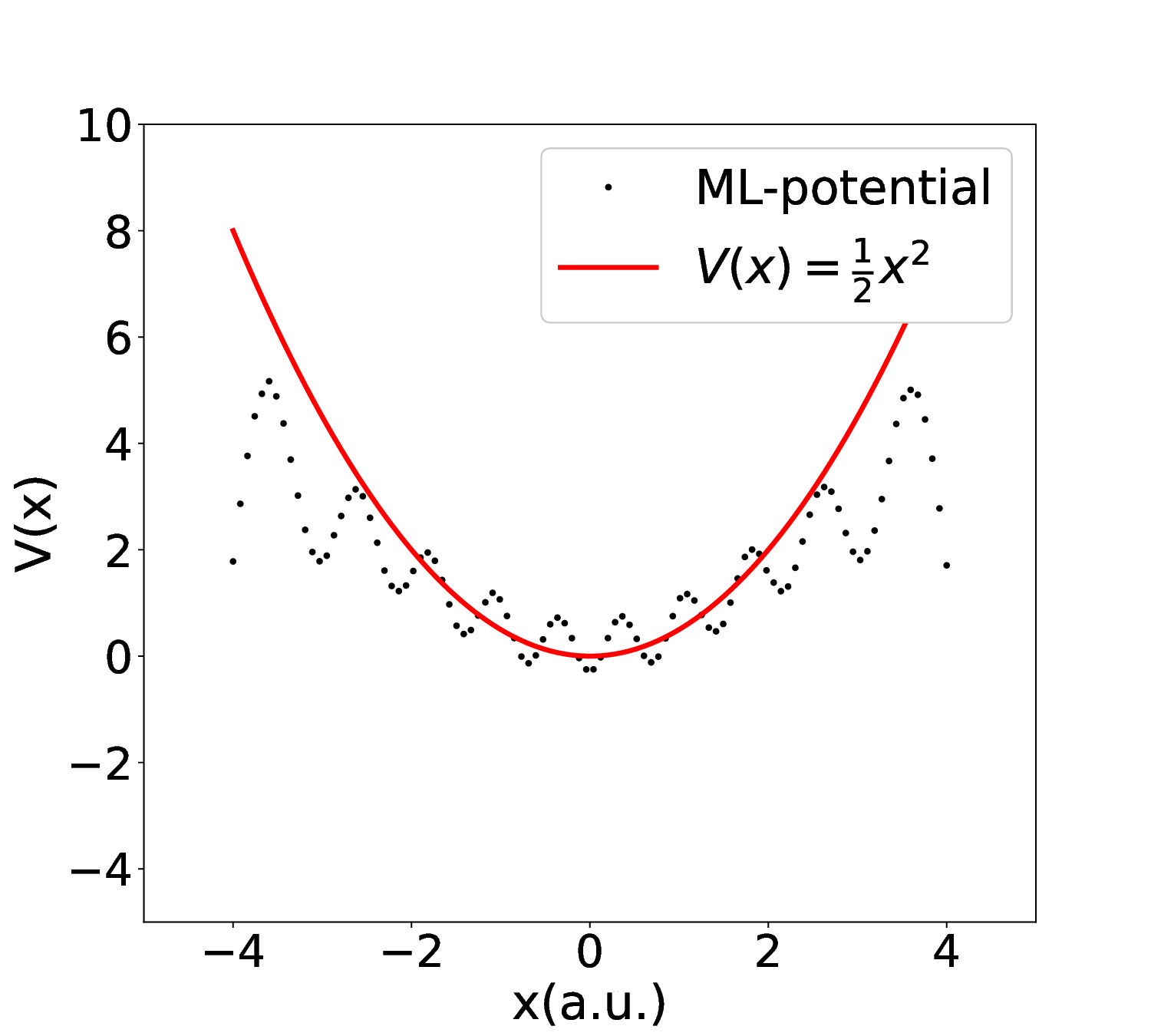}
    \caption{}
\end{subfigure}
\begin{subfigure}[b]{0.32\textwidth}
    \centering
    \includegraphics[width=\textwidth]{Images/V_15eigen_xmax6_150pts.eps}
    \caption{}
\end{subfigure}
\caption{Including more eigenstates in the training set increases the accuracy of the machine-learned potential. From left to right: (a) 5 eigenstates in the training set, (b) 5 eigenstates in the training set, (c) 5 eigenstates in the training set}
    \label{fig:comp}
\end{figure}

We show that having more eigenstates included in the dataset results in more accurate machine-learned Kohn-Sham potential. We examined this by training the neural network on three different datasets. Each of the datasets includes 5, 10, and 15 eigenstates. The results are shown in Fig.~\ref{fig:comp}. The three machine-learned Kohn-Sham potentials in the three figures share a similar trend. All of them capture the general feature of the actual Kohn-Sham potential and show deviations on the boundary of the system. The gap between the machine-learned potential and the exact potential on the boundary closes when more eigenstates are included.
\newpage
\section*{References}
\providecommand{\newblock}{}

\end{document}